\documentclass[prb,twocolumn,aps,notitlepage,amsmath,amsfonts]{revtex4-1}
\usepackage{amsmath,amsfonts,amssymb,graphicx,dsfont,bm,multirow,color}
\usepackage{soul}
\pdfoutput=1
\usepackage{hyperref}
\hypersetup{
    unicode=false,          
    pdftoolbar=false,        
    pdfmenubar=false,        
    pdffitwindow=false,     
    pdfstartview={FitH},    
    pdftitle={Symmetry breaking and Landau quantization in  topological crystalline insulators},    
    pdfauthor={M. Serbyn, and L. Fu},     
    pdfsubject={},   
    pdfcreator={},   
    pdfproducer={}, 
    pdfkeywords={} {} {}, 
    pdfnewwindow=true,      
    colorlinks=true,       
    linkcolor=[rgb]{0.09, 0.09, 0.5},          
    citecolor=[rgb]{0.09, 0.09, 0.5},        
    filecolor=magenta,      
    urlcolor=[rgb]{0.09, 0.09, 0.5}           
}
\def\be{\begin{eqnarray}}
\def\ee{\end{eqnarray}}
\def\bpm{\begin{pmatrix}}
\def\epm{\end{pmatrix}}

\newcommand{\bk}{{\bm k}}
\newcommand{\bp}{{\bm p}}

\newcommand{\DPpm}{E^\text{DP}_\text{L$\pm$}}

\newcommand{\DPhot}{E^\text{DP}_\text{H1/H2}}
\newcommand{\DPht}{E^\text{DP}_\text{H2}}

\newcommand{\Ld}{{\bm \Lambda}}
\newcommand{\bS}{{\bm S}}
\DeclareSymbolFont{matha}{OML}{txmi}{m}{it}
\DeclareMathSymbol{\varv}{\mathord}{matha}{118}
\begin{document}
\title{Symmetry breaking and Landau quantization in  topological crystalline insulators}
\author{Maksym Serbyn and Liang Fu}
\date{\today}
\begin{abstract} 
In the recently discovered topological crystalline insulators SnTe and Pb$_{1-x}$Sn$_{x}$(Te, Se), crystal symmetry and electronic topology intertwine to create topological surface states with many interesting features including Lifshitz transition, Van-Hove singularity and  fermion mass generation.  These surface states are  protected by mirror symmetry with respect to the (110) plane. In this work we present a comprehensive study of the effects of different mirror-symmetry-breaking perturbations on the (001) surface band structure. Pristine (001) surface states have four branches of Dirac fermions at low-energy. We show that ferroelectric-type structural distortion  generates a mass and gaps out some or all of these Dirac points, while strain shifts Dirac points in the Brillouin zone. An in-plane magnetic field leaves surface state gapless, but introduces asymmetry between Dirac points. Finally, an out-of-plane magnetic field leads to discrete Landau levels. We show that the Landau level spectrum has an unusual pattern of degeneracy and interesting features due to the unique underlying band structure. This suggests that Landau level spectroscopy can detect and distinguish between different mechanisms of symmetry breaking in topological crystalline insulators.
\end{abstract}
\affiliation{Department of Physics, Massachusetts Institute of
Technology, Cambridge, Massachusetts 02139}
\date{\today}

\pacs{
}

\maketitle

\section{Introduction}
The advent of topological insulators demonstrated the possibility for non-trivial band topology protected by time-reversal symmetry.~\cite{kanehasan, qizhang, moore} More recently, it was realized\cite{fu} that there exist topologically distinct classes of band structures that cannot be continuously deformed into each other without breaking certain crystal symmetries. Materials realizing such nontrivial band structures protected by crystal symmetry were termed topological crystalline insulators (TCI). The interplay between electronic topology and crystal symmetry dictates that TCI have gapless surface states on surfaces that preserve the corresponding crystal symmetry.  

The IV-VI semiconductor SnTe, as well as related alloys Pb$_{1-x}$Sn$_{x}$Te and Pb$_{1-x}$Sn$_{x}$Se,  were recently predicted\cite{hsieh} to belong to the TCI class protected by mirror symmetry\cite{teo} with respect to the (110) plane. This prediction was later verified by the direct observation of topological surface states in the ARPES experiments.~\cite{ando, story, xu} Signatures of surface states have also been observed in transport and scanning tunneling microscopy (STM)  measurements.~\cite{film, exp, yazdani} Remarkably, a recent STM experiment on (001) surface states in a magnetic field by Okada~\emph{et. al.}~\cite{exp} has found interesting features in the Landau levels that are not expected for a pristine TCI surface but consistent with a particular type of mirror symmetry breaking due to structural distortion\cite{hsieh}. 
This demonstrates the rich interplay between topology, crystal symmetry and electronic structure in topological crystalline insulators. 

In this paper, we present a comprehensive study of symmetry breaking in TCI surface states. We aim to understand the different ways in which the mirror symmetry breaking can be realized in a TCI and their effects on the surface band structure. Specifically, we consider the following three types of symmetry breaking: ferroelectric structural distortion, uniaxial strain and external magnetic field (or coupling to ferromagnetism). 

We study the effects of these perturbations on the (001) surface states of TCI, which exhibit various interesting features such as Lifshitz transition and Van-Hove singularity\cite{hsieh, JunweiLiu13, exp}.  For each type of perturbation, we use symmetry analysis to derive the form of its coupling to (001) surface states in $k\cdot p$ theory, and analyze its effect on the surface band structure. Our study of topological surface states under symmetry breaking provides the basic understanding of the band structure which is necessary to consider interaction effects. 

We further study Landau level (LL) spectrum of TCI surface states, which is a useful tool for detecting symmetry breakings.  We present a detailed calculation of LL spectrum of (001) surface states, and find many interesting features due to the unique band dispersion of TCI surface states. Our study of  LL spectrum is greatly needed for interpreting spectroscopic and transport measurements on TCI, and furthermore provides a starting point for studying interaction effects such as valley symmetry breaking and fractional quantum Hall effect. 

Our paper  is structured  as follows. In the next section we start with a brief summary of the four-band $k\cdot p$ model for the (001) surface states of TCI.  Section~\ref{Sec:3} is devoted to the effect of different mirror-symmetry breaking perturbations on the surface band structure. After this, in Section~\ref{Sec:4} we study the Landau level spectrum. Building upon understanding of the Landau levels without perturbations, we reveal how they are modified by different types of mirror  symmetry breakings. We conclude with the summary of the main results in Section~\ref{Sec:5}.

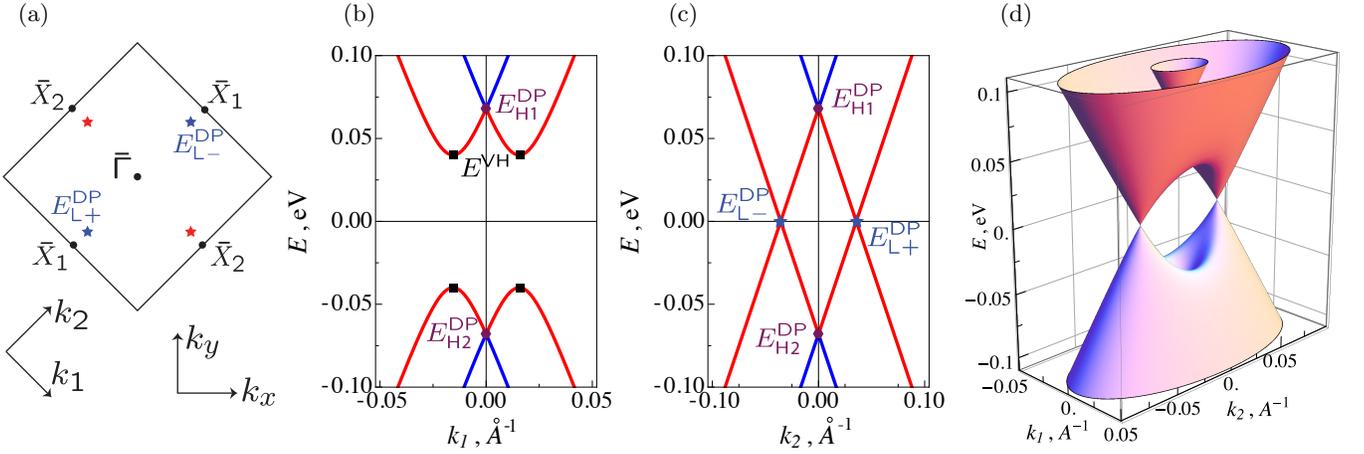
\begin{figure*}
\begin{center}
\includegraphics[width=0.44\columnwidth]{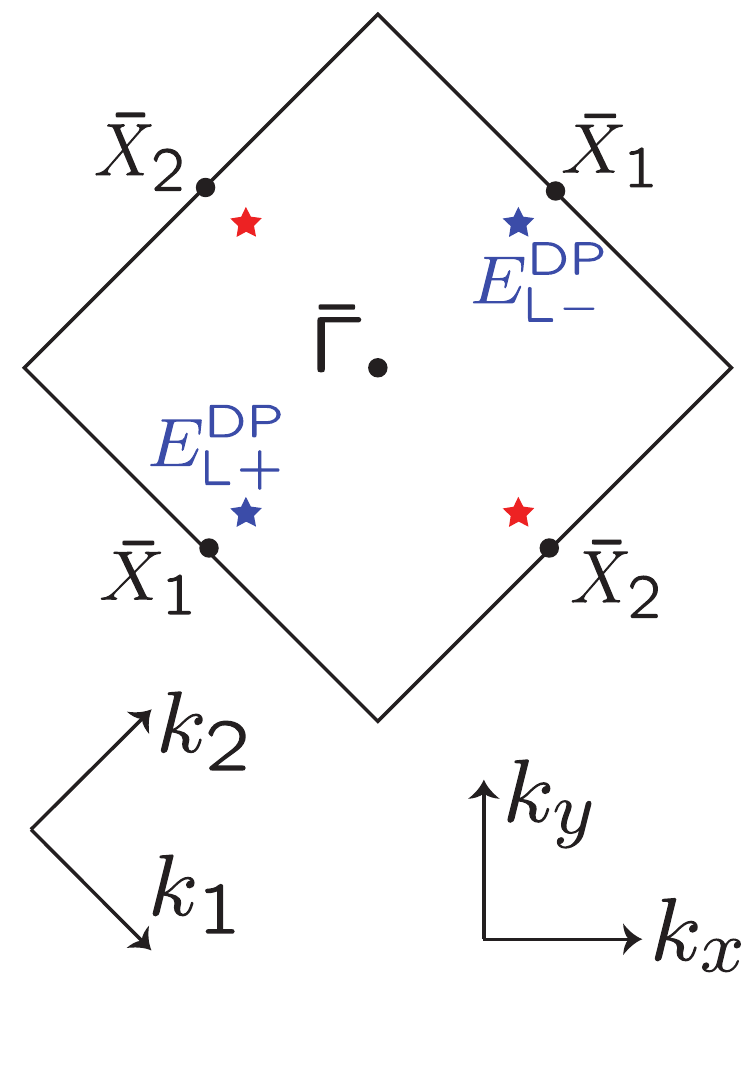}
\includegraphics[width=0.5\columnwidth]{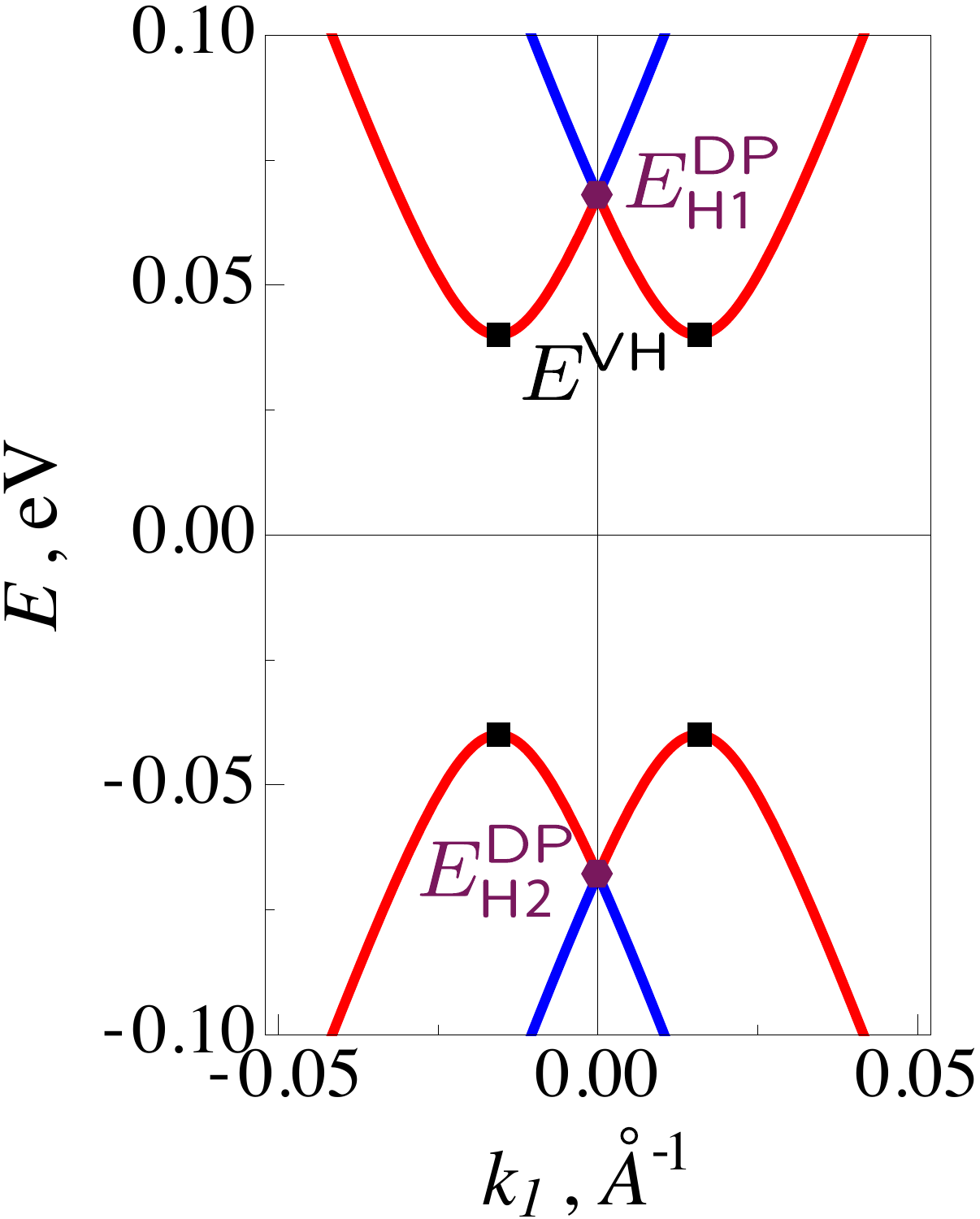}
\includegraphics[width=0.5\columnwidth]{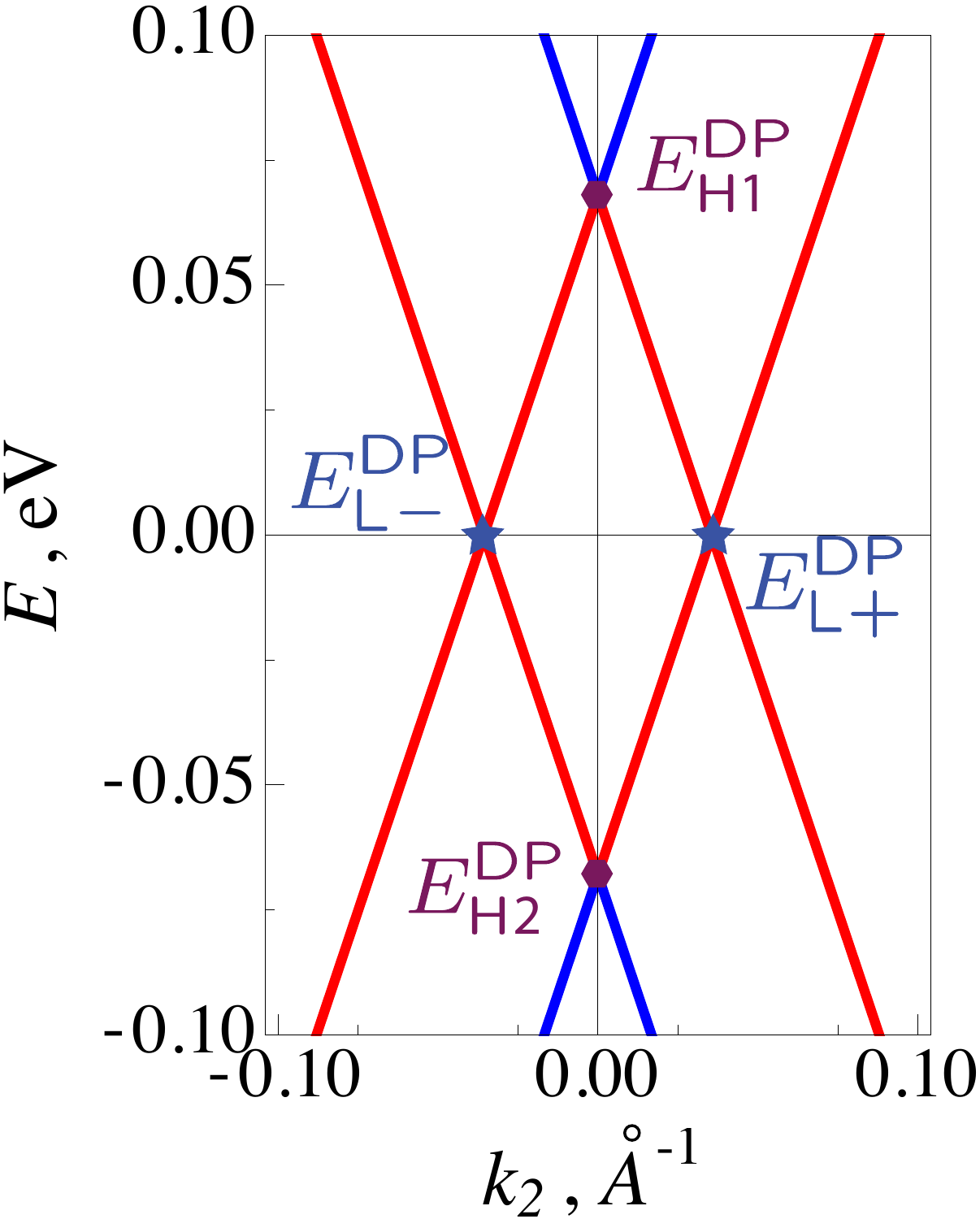}
\includegraphics[width=0.6\columnwidth]{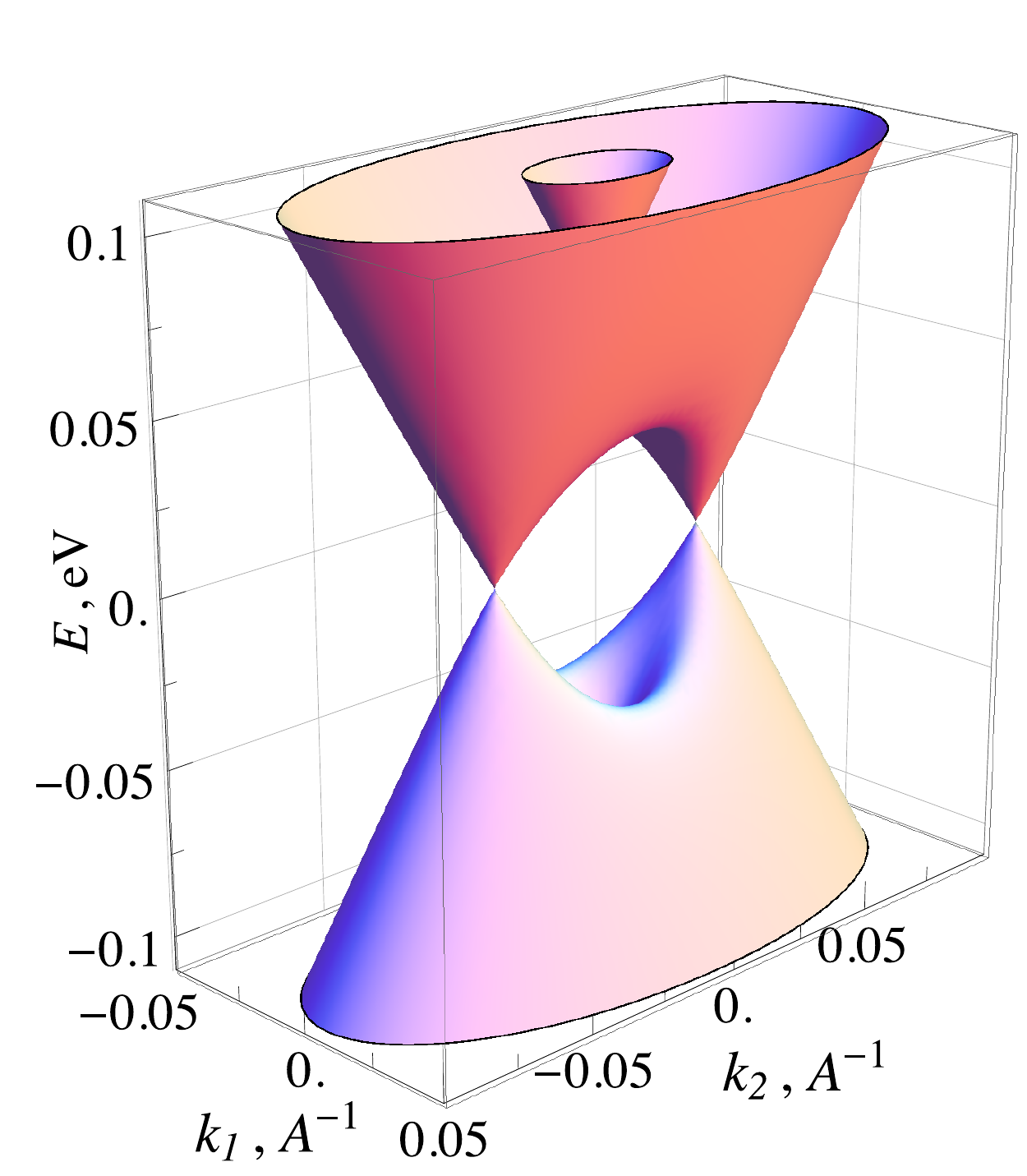}
\setlength{\unitlength}{\columnwidth}
\begin{picture}(2,0)
\put(0.0, 0.7){(a)}
\put(0.5, 0.7){(b)}
\put(1., 0.7){(c)}
\put(1.51, 0.7){(d)}
\end{picture}
\caption{ \label{F:band} (a) $\bar X_{1,2}$ points within the surface Brillouin zone.  $k_{x,y}$ correspond to the standard coordinates, whereas $k_{1,2}$ are the coordinates adopted in the paper. (b-d) Emergent  band structure in vicinity of $\bar X_1$ point has two low-energy Dirac cones which merge at higher energies. Parameters used are $v_1 = 3.53~\text{eV}\cdot\text{\AA}$, $v_2 = 1.91~\text{eV}\cdot\text{\AA}$, and $m = 0.055~\text{eV}$, $\delta = 0.04~\text{eV}$ taken from  Ref.~\onlinecite{exp}.}
\end{center}
\end{figure*}

\section{$k\cdot p$ model of surface states} 

\subsection{Four-band model}
We start by reviewing the four-band $k\cdot p$ model for the (001) surface states of TCI derived in Ref.~\onlinecite{JunweiLiu13}, which captures all essential features of the (001) surface states.   
This model is directly derived from the bulk Dirac fermion band structure of TCI, with inverted Dirac mass, in the vicinity of four distinct $L$ points in the three-dimensional Brillouin zone~(BZ).  
For the (001) surface,  two out of the four $L$ points, $L_{1,2}$ are projected to the $\bar X_1$ and the remaining two projected to $\bar X_2$ points of the surface BZ, see Fig.~\ref{F:band}(a). To derive the (001) surface states, as a first step we consider a interface between SnTe, a representative TCI, and its non-topological cousin PbTe, which is topologically equivalent to the vacuum.  The low-energy band structures of both SnTe and PbTe can be modeled by three-dimensional Dirac fermions, with masses of opposite signs.  
Therefore a {\it smooth} interface between SnTe and PbTe hosts two-dimensional massless Dirac fermions,\cite{Volkov,Fradkin86}  known as domain wall fermions. Importantly, among the four $L$-valleys, both $L_1$ and $L_2$ ($L_3$ and $L_4$) are projected to the $X_1$ ($X_2$) point on the (001) surface Brillouin zone. This leads to two {\it degenerate} branches of two-dimensional massless Dirac fermions at $X_1$ ($X_2$), described by the effective Hamiltonian
\begin{equation} \label{Eq:H0}
H_{{\bar X}_1}^0(\bk) =  (v_1 k_1  s_y - v_2 k_2 s_x) \otimes \tau_0,  
\end{equation}
where $\bk$ is measured from the $\bar X_1$ ($\bar X_2$) point, see Fig.~\ref{F:band}(a). 
The Pauli matrices $\vec s = (s_x,s_y,s_z)$ act in the space of Kramers doublet, 
and $\vec \tau = (\tau_x,\tau_y,\tau_z)$ act in the valley space $L_1$ and $L_2$. 
$\tau_0$ is the identity matrix, and for simplicity of the notation, tensor product with $\tau_0$ will be omitted in what follows.

The real SnTe (001) surface states are rather different from the above domain wall fermions. 
Because of the atomically sharp boundary between SnTe and the vacuum, scattering between $L_1$ and $L_2$ valleys, and between $L_3$ and $L_4$ valleys, is present at the SnTe (001) surface.  As first shown in Ref.~\onlinecite{JunweiLiu13}, this inter-valley scattering hybridizes the  two degenerate interface Dirac fermions to create the actual (001) surface states of TCI. For the sake of completeness, we review the derivation of Ref.~\onlinecite{JunweiLiu13} below. 

To capture the inter-valley hybridization, we introduce off-diagonal terms in the valley basis into the $k\cdot p$ model, 
which must respect all the symmetries of the (001) surface. 
There are three crystal symmetries that leave the $\bar X_1$ point invariant: the two-fold rotation around surface normal, $C_2$, as well as two independent mirror symmetries with respect to reflection of $x_1$ and $x_2$ axes, $M_{1}: x_1 \to -x_1$ and $M_{2}: x_2 \to -x_2$. Also, the time-reversal symmetry, denoted as $\Theta$, is present. The action of these symmetry operations are represented by corresponding $4 \times 4$ operators in the $k \cdot p$ model that act on spin/valley space.  
Specially, mirror reflection $M_1$ acts  on the electron's spin but leaves each valley intact, while the mirror $M_2$ and the two-fold rotation $C_2$ interchanges $L_{1}$ and $L_2$, in addition to acting on the spin. Therefore these symmetries are represented as follows:
\begin{subequations}\label{Eq:sym}
\begin{eqnarray} 
&C_2&: - i \tau_x s_z ,\\
&M_1&: - i s_x  ,\\
&M_2&: - i \tau_x s_y,  \\
&\Theta&: i s_y K,
\end{eqnarray}
\end{subequations}
where $K$ is the complex conjugation. 

There exist only two additional lowest-order terms which are invariant under all symmetries listed in Eq.~(\ref{Eq:sym}). These are   $\tau_x$ and $\tau_y s_x$ (see Table~\ref{Tab:symm}): they arise from valley hybridization that occurs at  the atomically sharp surface.  Adding these terms to the Hamiltonian~(\ref{Eq:H0}), we  obtain the effective Hamiltonian for the TCI (001) surface states:~\cite{JunweiLiu13}
\begin{equation}\label{Eq:Hsurf}
H_{{\bar X}_1}(\bk) =  v_1 k_1   s_y - v_2 k_2 s_x  + m  \tau_x + \delta s_x    \tau_y. 
\end{equation}
With four parameters, Eq.~(\ref{Eq:Hsurf}) is our starting point in studying the effect of symmetry breaking perturbations on TCI (001) surface states. 
Additional corrections~\cite{fang, WangPRB13}  to $H_{{\bar X}_1}(\bk)$, which are proportional to $\bk$   are unimportant,~\cite{JunweiLiu13} and will not be considered below.  

The dispersion of the four-band Hamiltonian~(\ref{Eq:Hsurf}) can be visualized starting from two degenerate Dirac cones described by Eq.~(\ref{Eq:H0}). The Dirac points of these cones initially are located precisely at the $\bar X_1$ point in the momentum space and at zero energy.  
The first intervalley term $m\tau_x$ shifts the energy of two Dirac cones from zero to positive and negative energies  $E^\text{DP}_\text{H$1$} = +m$ and 
$E^\text{DP}_\text{H$2$} = -m$. The upper (lower) Dirac point is mainly derived from the Te (Sn) $p$-orbitals.~\cite{Safaei13}
The two-components of each Dirac point form a Kramers doublet at $X_1$. These upper and lower Dirac cones are hereafter referred to as ``high-energy'' Dirac cones.  

For $\delta=0$, the lower Dirac cone associated with  $E^\text{DP}_\text{H$1$}$ overlaps with the upper Dirac cone associated with $E^\text{DP}_\text{H$2$}$ on a  ring in $\bk$-space at zero energy. A nonzero second invervalley term $\delta s_x \tau_y$  in~(\ref{Eq:Hsurf}) lifts this degeneracy everywhere except for two points on the axis $k_1 = 0$, where  two bands with opposite mirror eigenvalues $\pm i$ (associated with the reflection $M_1$) cross each other. This anisotropic band hybridization generates a pair of Dirac points at energy $E=0$, which are located on opposite sides of $\bar{X}_1$ at momenta
\begin{equation} \label{Eq:DiracCone}
\Ld_\text{$\pm$} = (0, \pm \sqrt{m^2+\delta^2}/v_2)
\end{equation}
measured from the $X_1$ point, see Fig.~\ref{F:band}~(c-d). These two Dirac points are descendants of the high-energy Dirac points and will be referred to as ``low-energy''. 

In addition to generating the low-energy Dirac cone, the $\delta$ term further pushes the high-energy Dirac points apart from each other by level repulsion. The renormalized Dirac point energies are given by  
\begin{equation} \label{Eq:highDirac}
E^\text{DP}_\text{H1 (H2)}
=
\pm\sqrt{m^2+\delta^2}.
\end{equation}

Last but not the least,  the above anisotropic band hybridization described by the $\delta$ term generates a pair of saddle points in the band dispersion near $X_1$\cite{hsieh, JunweiLiu13}, which are located at an intermediate energy $E^{\text{S}}$ on the line $\Gamma X_1$ with momenta $\bS_\pm$~[see Fig.~\ref{F:band}~(b)]: 
\begin{equation} \label{Eq:VHS}
E^{\text{VH}} =\delta, \; \; \bS_\pm= \left(\pm m/v_1, 0\right).
\end{equation}
At $E^{\text{VH}}$, the density of states diverges, leading to a Van-Hove singularity (VHS). 
Another pair of saddle points exist at the negative energy $-E^{\text{VH}}$.  These saddle points are associated with a change of Fermi surface topology as a function of Fermi energy, i.e., Lifshitz transition. 
For energies below Van-Hove singularity, $|E|<E^{\text{VH}}$, the Fermi surface consists of two disconnected pockets of the two low-energy Dirac cones. Above $E^{\text{VH}}$, these two pockets merge into two concentric ellipses with opposite types of carriers, which are centered at $X_1$ and associated with the high-energy Dirac cones.

\subsection{Two-band model}

In what follows we will be mainly concerned about low-energy properties and their modification upon addition of  weak symmetry breaking perturbations. Therefore, it is convenient to  linearize the band structure of the four-band model (\ref{Eq:Hsurf}) near $\Ld_\pm$ and obtain a two-band model for the low-energy Dirac fermions. 
Let us first consider the Dirac cone at $\Ld_+$. 
Introducing a new set of Pauli matrices $\vec \mu=(\mu_x,\mu_y,\mu_z)$ for the two degenerate states at $\Ld_+$ and projecting (\ref{Eq:Hsurf}) onto the corresponding subspace, we obtain 
the desired two-band Hamiltonian\cite{hsieh, JunweiLiu13}:   
\begin{equation} \label{Eq:H0-proj}
H_{\Ld_+}(\bp)
  =
 v'_1 p_1 \mu_y -  v_2 p_2 \mu_x, 
\end{equation}
where the momentum $\bp$ is measured from $\Ld_+$, and 
$ v_1'= \frac{\delta}{\sqrt{m^2+\delta^2}} v_1$. 
It should be noted that the two components of this low-energy Dirac point, $\mu_x=\pm 1$, correspond to Te and Sn $p$-orbitals respectively,~\cite{Safaei13} which are {\it not} Kramers doublet. The two-band Hamiltonian for the other Dirac cone at $\Ld_-$, $H_{\Ld_-}(\bp)$, is simply related to $H_{\Ld_+}(\bp)$ by the two-fold rotation $C_2$. 

So far we have been describing the band structure in vicinity of the $\bar X_1$ point in the BZ. In the absence of symmetry breaking perturbations, the $\bar X_1$ and $\bar X_2$ points are related to each other by a rotation of $\pi/2$, so that the band structure near $\bar{X}_1$ has a symmetry-related copy near $\bar X_2$ point. As a consequence, we can deduce   the effect of the perturbations on the $\bar X_2$ point from that on the  $\bar X_1$ point  by symmetry considerations. For example, the effect of a magnetic field $B_1$, parallel to $k_1$ axis on the $\bar X_2$ point can be deduced from the effect of magnetic field $B_2$, parallel to $k_2$ axis on the $\bar X_1$. For this reason, in the rest of this work we will explicitly consider the $\bar X_1$ point only.

\section{Mirror symmetry breaking \label{Sec:3}}

We now analyze the effects of various symmetry-breaking perturbations. Since mirror symmetry is crucial for defining the electronic topology in the SnTe class of TCI, 
one might expect that an infinitesimal  mirror symmetry breaking  is sufficient to open up a gap for TCI surface states. However, we find this is not always the case. Instead, different mirror-symmetry breaking perturbations act differently on the (001) TCI surface states, depending on other symmetry properties. Our findings have significant implications for new classes of topological crystalline insulators that are protected by other crystal symmetries, which we will reveal in Section~\ref{Sec32} below.

\begin{figure}[t]
\begin{center}
\includegraphics[width=1.\columnwidth]{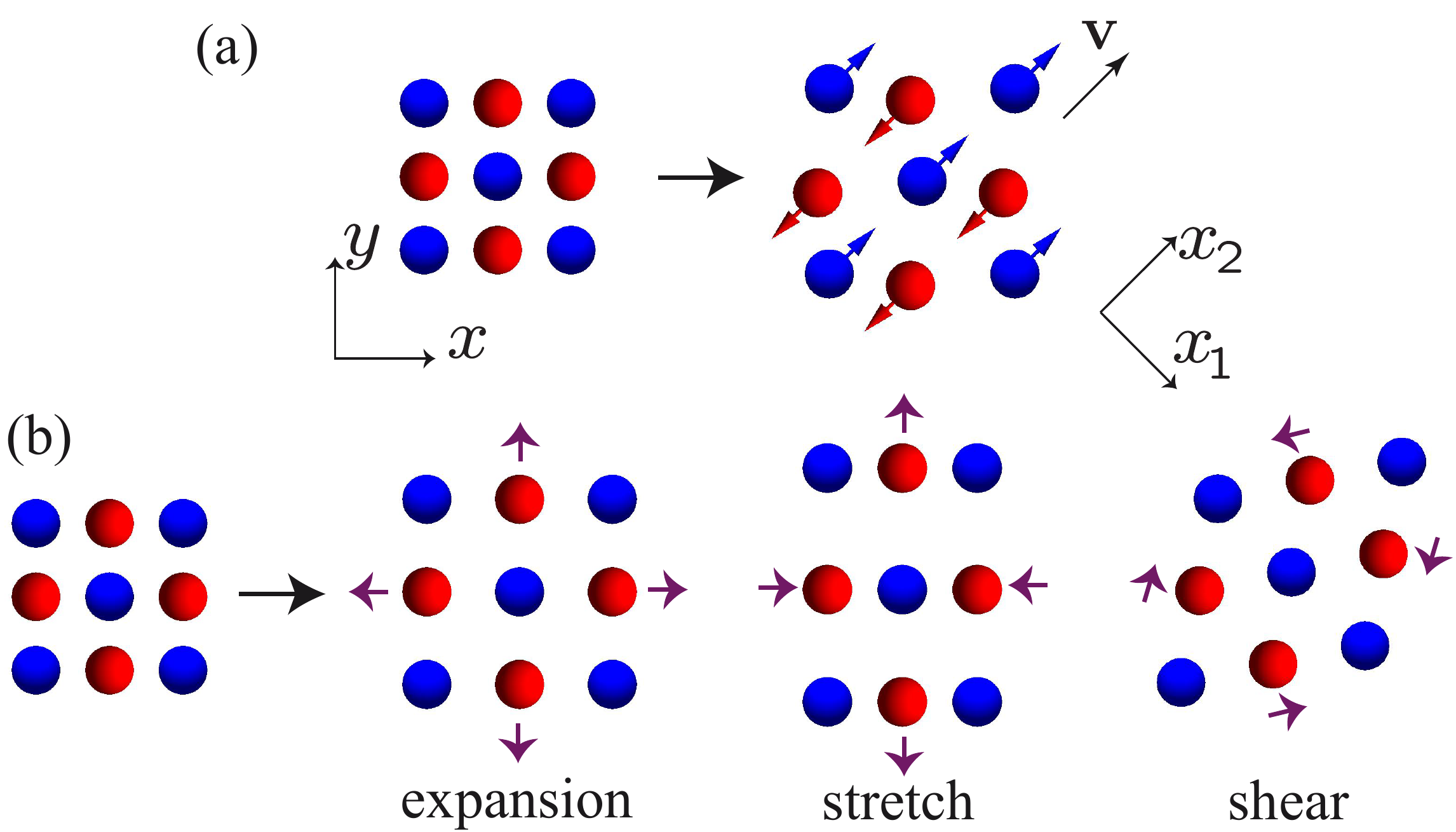}
\caption{ \label{F:symm} (a) Cartoon picture of ferroelectric distortion, which displaces two kinds of atoms with respect to each other. The corresponding order parameter, denoted by $\mathbf v$ is parallel to [110] direction ($x_2$ axis), and breaks only $M_2$ mirror plane. (b) Schematic of decomposition of generic strain into three components  corresponding to expansion, uniaxial strain or stretch, and shear, each with different symmetry properties. }
\end{center}
\end{figure}

In this work, we consider the following three common types of perturbations to TCI in the SnTe material class.  
 
 \begin{enumerate}
 \item[(i)] Structural distortion. This corresponds to a displacement of Sn and Te atoms along opposite directions, $\mathbf{v}$ and $-\mathbf{v}$, which occurs spontaneously  in SnTe at low temperature.~\cite{Brillson74} This distortion fully breaks the rotation symmetry see Fig.~\ref{F:symm}(a), and leads to a nonzero ferroelectric polarization.  A ferroelectric displacement with $\mathbf{v}=(\varv_1, 0)$ parallel to the $x_1$ axis breaks one mirror symmetry $M_1$, but is invariant under $M_2$; and vice versa for $\mathbf{v}=(0, \varv_2)$. 

\item[(ii)] Strain. Generic strain can be decomposed into expansion, stretch and shear with different symmetry properties as shown in Fig.~\ref{F:symm}(b). Stretch deformation  has the most interesting effect, since it breaks both mirror symmetries $M_1$ and $M_2$, as well as the rotation $C_4$, while preserving $C_2$. 

\item[(iii)] Zeeman coupling to either external magnetic field or ferromagnetic moment in magnetically doped TCI, e.g., Sn$_{1-x}$Mn$_x$Te and Sn$_{1-x}$Eu$_x$Te. An in-plane Zeeman field  fully breaks the rotation symmetry, but is invariant under the combined operation of two-fold rotation and time-reversal, while a perpendicular field preserves the rotation symmetry. 
Moreover, since magnetic field is a {\it pseudovector}, it transforms under mirror symmetry in the opposite way to the ferroelectric displacement vector. 
Specifically, an out-of-plane field $B_3$ breaks both mirror symmetries, while 
an in-plane field $B_1$ parallel to the $x_1$ axis preserves the mirror symmetry $M_1$ ($x_1 \rightarrow -x_1$), but breaks the mirror symmetry $M_2$ ($x_2 \rightarrow -x_2$), and vice versa for $B_2$.   

\end{enumerate}

 \begin{table}[t]
\begin{center}
\begin{tabular}{lccccccc c c }
Perturbation&$\Theta$&$M_1$ &$M_2$ & $C_2$ & Matrices&  Effect on $\DPpm$\\
\hline
\hline
Symmetry preserving & + &+ &+ &+ & $\tau_x$; $s_x\tau_y$& shift $ \Ld$ \\
\hline
Ferroelectric dist. $\mathbf{v}||(1\bar10)$& + &+ &$-$ &$-$ & $\tau_z$ & open gap\\
\hline
Ferroelectric dist. $\mathbf{v}||(110)$& + &$-$ &+ &$-$ & $s_z\tau_y$ & ---\\
\hline
Uniaxial strain $u_{xx}-u_{yy}$ & + &$-$ &$-$ &$+$ & $s_y\tau_y$&shift $ \Ld$ \\
\hline
\multirow{3}{*}{Magnetic field $\mathbf{B}||(1\bar10)$} & \multirow{3}{*}{$-$} &\multirow{3}{*}{$+$} &\multirow{3}{*}{$-$} &\multirow{3}{*}{$-$}& $s_x$&\ shift $\Lambda_2$$^*$ \\
& &&&& $\tau_y$& shift $E^\text{DP}$ \\
&&&&&  $s_x\tau_x$  & shift $E^\text{DP}$ \\
\hline
\multirow{3}{*}{Magnetic field $\mathbf{ B}||(110)$} & \multirow{3}{*}{$-$} &\multirow{3}{*}{$-$} &\multirow{3}{*}{$+$} &\multirow{3}{*}{$-$} & $s_y$& shift $\Lambda_1$$^*$ \\
& &&&& $s_y\tau_x$  &shift $\Lambda_1$\\
& &&&&$s_z \tau_z$& shift $ \Lambda_1$$^*$ \\
\hline
\multirow{3}{*}{Magnetic field $\mathbf{ B}||(001)$} & \multirow{3}{*}{$-$} &\multirow{3}{*}{$-$} &\multirow{3}{*}{$-$} &\multirow{3}{*}{$+$}& $s_z$& open gap$^*$ \\
&&&&&  $s_y\tau_z$   & open gap$^*$\\
& &&&& $s_z\tau_x$ &  --- \\
\hline\hline
\end{tabular}
\end{center}
\caption{\label{Tab:symm} Effects of different perturbations on TCI (001) surface states, classified by their symmetry properties. 
The plus (minus) sign indicates that the perturbation is even (odd) under the corresponding symmetry transformation.  The form of each perturbation in the four-band model (\ref{Eq:Hsurf}) is shown explicitly as a $4\times 4$ matrix in valley ($\tau$) and spin ($s$) space, or matrices if there is more than one.  
The effect of the perturbation on low-energy Dirac points is described in the last column: symmetry breaking perturbations can either open gaps at Dirac points, or shift their positions in momentum space ($\Ld$) or change their energies ($E^\text{DP}$).  Presence (Absence) of the asterisk in the last column indicates that the effect is of the same (opposite) sign for the two Dirac points $\DPpm$.
}
\end{table}%

Table~\ref{Tab:symm} summarizes our main results, showing the symmetry properties of these perturbations  (columns II--V), their explicit forms in the four-band Hamiltonian (column VI), and their effects on the low-energy Dirac fermions on the TCI (001) surface (last column).  Some of these perturbations have been considered\cite{hsieh, fang} using the phenomenological two-band Hamiltonian only.  In contrast, our results are derived from the full four-band theory and thus capture the effects of perturbations in the whole energy range of TCI surface states.  

Our derivation is based entirely on symmetry analysis. Specifically, based on the symmetry transformations (\ref{Eq:sym}),  we enumerate all lowest-order terms that transform in the same manner as the perturbation under consideration. For example, the ferroelectric distortion ${\mathbf v} = (\varv_1, 0)$ must couple uniquely to the operator $s_z \tau_y$, because both are even under time-reversal and $M_2$, and odd under $M_1$ and $C_2$. By carrying out similar analysis for all other perturbations, we derive their forms in four-band Hamiltonian, as listed in column VI of Table I.  After this, we project these perturbations from the four-band Hamiltonian to the two-band Hamiltonian that describes the low-energy Dirac cone at $\Ld_+$, and list their effect in the last column of Table I (corresponding terms in the low-energy Hamiltonian are listed below). In what follows we discuss the effect of each perturbation in more details.  
 
 \subsection{Ferroelectric distortion \label{Sec31}}

As explained above, symmetry analysis dictates that ferroelectric displacements in the $(110)$ and $(1\bar{1}0)$ direction, $\varv_1$ and $\varv_2$, couple to the surface states near $\bar{X}_1$ in the following form:
\begin{equation}\label{HM21}
V_F = g_{F1} \varv_1 s_z\tau_y+  g_{F2} \varv_2 \tau_z,
\end{equation}
where $g_{Fj}$ parametrizes the coupling strengths. 

The two terms in Eq.~(\ref{HM21}) have dramatically different effects on the surface states near $\bar{X}_1$. 
$\varv_1$ breaks the mirror symmetry $M_1$ that protects the low-energy Dirac point $\Ld_\pm$, and hence opens up a band gap there. 
This is verified by projecting onto low energy Hilbert space at the Dirac point $\Ld_+$: we find the two-band Hamiltonian is given by
 \begin{equation} \label{Eq:HM1proj}
\tilde{V}_{F} = \Delta_F \mu_z,
\end{equation}
i.e., the mirror symmetry breaking generates a Dirac mass $\Delta_F \propto \varv_1$ at $\Ld_+$, and thus opens up a gap $E_{g, \Ld_+} = | \Delta_F|$.  

It follows from time-reversal symmetry that the above distortion also generates a gap $E_{g, \Ld_-} = E_{g, \Ld_+}$ at the other Dirac point $\Ld_-$. 
However, the sign of the Dirac mass at $\Ld_-$ remains to be determined. 
Throughout this work, we adopt the convention\cite{hsieh} that Dirac masses at $\Ld_+$ and $\Ld_-$ are equal if the two Dirac points are related by the two-fold rotation $C_2$. However, the ferroelectric distortion considered here breaks $C_2$, so that the resulting Dirac mass at $\Ld_-$ is $- \Delta_F$, opposite to the one at $\Ld_+$.   

Unlike the ferroelectric distortion $\varv_1$, the $\varv_2$ term in (\ref{HM21}) vanishes when projected onto the low-energy Dirac points. This is consistent with the fact that non-zero component $\varv_2$ does not break the mirror symmetry $M_1$ which protects the massless Dirac fermions at $\Ld_\pm$. 

\subsection{Strain \label{Sec32}}

Generic strain, described by a displacement field $\bm u$ can be represented as a superposition of uniform expansion, uniaxial strain (or stretch) which conserves volume and a shear deformation. All three of these are schematically depicted in Fig.~\ref{F:symm} (b). More formally, in the coordinate system coinciding with a principal crystal axes, the uniform expansion is represented as $\partial_x u_{x}+\partial_y u_{y}$, or $u_{xx}+u_{yy}$ in the short hand notations. Such combination is invariant under all symmetries, and thus it can only change the parameters $m$ and $\delta$ in the four-band Hamiltonian~(\ref{Eq:Hsurf}). This causes a shift in the position of the low-energy Dirac points $\Ld_+$  and $\Ld_-$ along the mirror-symmetric line $\Gamma \bar{X}_1$ in opposite directions by an equal amount. This Dirac point shift under uniform strain, which we deduce from symmetry analysis here,  has been found in recent ab-initio calculations.~\cite{Barone13,Quian-unp} 

The shear deformation $u_{xy}+u_{yx}$ breaks only $C_4$ rotation symmetry, but respects $M_1$, $M_2$ mirror planes as well as $C_2$ rotation. Therefore, the shear can change parameters $m$ and $\delta$ in a different way in vicinity of $\bar X_1$ and $\bar X_2$ points, but do not induce any new terms. 
 
The most interesting case is the uniaxial strain, written as $u_{xx}-u_{yy}$, which breaks  mirror symmetries $M_1$ and $M_2$ as well as $C_4$, but preserves $C_2$ and time-reversal symmetry. Despite this symmetry breaking, we find that the uniaxial strain deformation does not open a gap in the low-energy Dirac cones. The gapless nature of Dirac points is protected by the rotation symmetry $C_2$ in combination with time-reversal symmetry, denoted by $\Xi \equiv \Theta C_2$. $\Xi$  is an anti-unitary operator satisfying $\Xi^2 =1$, and thereby imposes a reality condition on the surface state wavefunction at every momentum. This leads to a quantized $\pi$ Berry phase that protects the gapless Dirac point, while the position of the low-energy Dirac point can be shifted in both $k_1$ and $k_2$ directions by the stretch deformation. Furthermore, time-reversal symmetry dictates that  the two Dirac points, $\Ld_+$ and $\Ld_-$, shift in opposite directions by an equal amount.    

The presence of such Dirac cone located at a completely {\it generic} momentum signals a new class of topological crystalline insulators protected by two-fold rotation and time-reversal symmetry, instead of mirror symmetry. This interesting subject will be described elsewhere.    

\subsection{In-plane magnetic field \label{Sec33}}

We now study the effect of an in-plane magnetic field, with two components $B_1$ and $B_2$.
Based on symmetry analysis, we find the following allowed coupling terms  in the four-band Hamiltonian:
\begin{subequations}\label{Eq:HB12}
\begin{eqnarray} \label{Eq:HB1}
V_{B_1}
&=&
 \mu^B_{1} B_1 s_x+
\eta_{1} B_1 \tau_y+
 \lambda_{1} B_1 s_x\tau_x,
 \\ \label{Eq:HB2}
V_{B_2}
&=&
 \mu^B_{2}  B_2 s_y
 +
 \eta_{2}B_2 s_y\tau_x
 +
  \lambda_{2} B_2 s_z\tau_z.
\end{eqnarray}
\end{subequations}

To analyze the effect of an in-plane magnetic field on TCI surface states, we project $V_{B_1}$ and $V_{B_2}$ onto the low-energy subspace associated with the Dirac cone at $\Ld_+$. 
We find that the leading effect of the in-plane field, given by the terms proportional to $\mu^B_{1,2}$ in~(\ref{Eq:HB12}), is to \emph{shift} the position of the Dirac cones in BZ. 
For $B_1 \neq 0$ and $B_2 =0$ ($B_1=0$ and $B_2 \neq 0$), the Dirac point $\Ld_+$ near $X_1$ shits along the $k_2$ ($k_1$) direction, in agreement with the fact that magnetic field is a pseudo-vector and thus a nonzero $B_1$ ($B_2$) preserves the mirror symmetry $M_1$ ($M_2$). The Dirac point $\Ld_-$ shifts along the opposite direction.  

Terms proportional to $\eta_{1,2}$ and $\lambda_{1,2}$ in Eq.~(\ref{Eq:HB12}) arise from inter-valley mixing at the surface and thus are expected to be subleading. Nevertheless, we briefly mention their effect. The last two terms in~(\ref{Eq:HB1}) shift the energy of the low-energy Dirac points $\DPpm$ away from zero by an amount 
\begin{equation}
\Delta E = - \frac{{\delta} \eta_1 + m \lambda_1 }{\sqrt{m^2+\delta^2}} B_1.
\end{equation}
On the other hand, the remaining two terms in Eq.~(\ref{Eq:HB2})  shift the position of the Dirac points within the BZ, $\Ld_\pm$.

For $B_1 \neq 0$ and $B_2 \neq 0$, both $M_1$ and $M_2$ symmetries are broken. This causes the Dirac points to shift their locations and energies, but does not generate any gap.  Similar to the case of uniaxial strain~[see Section~\ref{Sec32}], each gapless Dirac cone is now located at a generic momentum, and it is protected by the {\it combination} of $C_2$ and  time-reversal symmetry, which remains intact in the presence of an in-plane field. This signals a new class of topological crystalline insulators protected by the  symmetry $\Xi = \Theta C_2 $. We note that the combination of time-reversal and lattice translation symmetries could also lead to topological phases such as antiferromagnetic topological insulator,~\cite{Mong10}  see also Ref.~\onlinecite{Liu13}.

\subsection{Perpendicular magnetic field \label{Sec34}}
In contrary to the in-plane magnetic field which has only Zeeman-type couplings to the surface states,  the perpendicular magnetic field leads to appearance of Landau levels. We postpone the discussion of the Landau levels spectrum until next section, and concentrate on the allowed Zeeman-like couplings and their effect. Such a Zeeman-only effect can also 
arise from exchange interaction between conduction electrons and localized moments in magnetically doped TCI. 

From symmetry analysis we deduce the following form of Zeeman coupling of TCI surface states to a perpendicular magnetic field or magnetic moment:
\begin{equation} 
\label{Eq:B3}
 V_{B_3}
=
\mu^B_{3} s_z
+
\eta_{3} s_y\tau_z
+
\lambda_{3} s_z\tau_x.
\end{equation}
Projection of Eq.~(\ref{Eq:B3}) onto the low-energy Dirac cone $\Ld_+$ generates a Dirac mass 
\begin{eqnarray} \label{Eq:projBall}
V_{B_3} = m_{B_3} \mu_z,  
\end{eqnarray}
where $m_{B_3} =-  ({\delta}/{\sqrt{m^2+\delta^2}})\mu^B_3 -   ({m}/{\sqrt{m^2+\delta^2}})\eta_3$. 

In contrary to the mass generated by ferroelectric distortion~[Eq.~(\ref{Eq:HM1proj})], which has opposite signs for two nearby Dirac cones, in this case the mass is of the same sign for both Dirac points. This difference leads to a remarkable consequence: the TCI (001) surface with Zeeman gap realizes a two-dimensional quantum anomalous Hall (QAH) state with quantized Hall conductance $\sigma_{xy}=2 \mathop{\rm sign}(B_3) {e^2}/{h}$, as shown in Ref.~\onlinecite{hsieh}. In a TCI (001) thin film,  the  top and bottom surfaces add up to form a QAH state with $\sigma_{xy}=\pm 4  {e^2}/{h}$ (see Ref.~\onlinecite{Bernevig13}), provided that the hybridization between the two surfaces are relatively weak.~\cite{LiuNatMat13}

 \begin{figure*}
\begin{center}
\includegraphics[width=0.49\columnwidth]{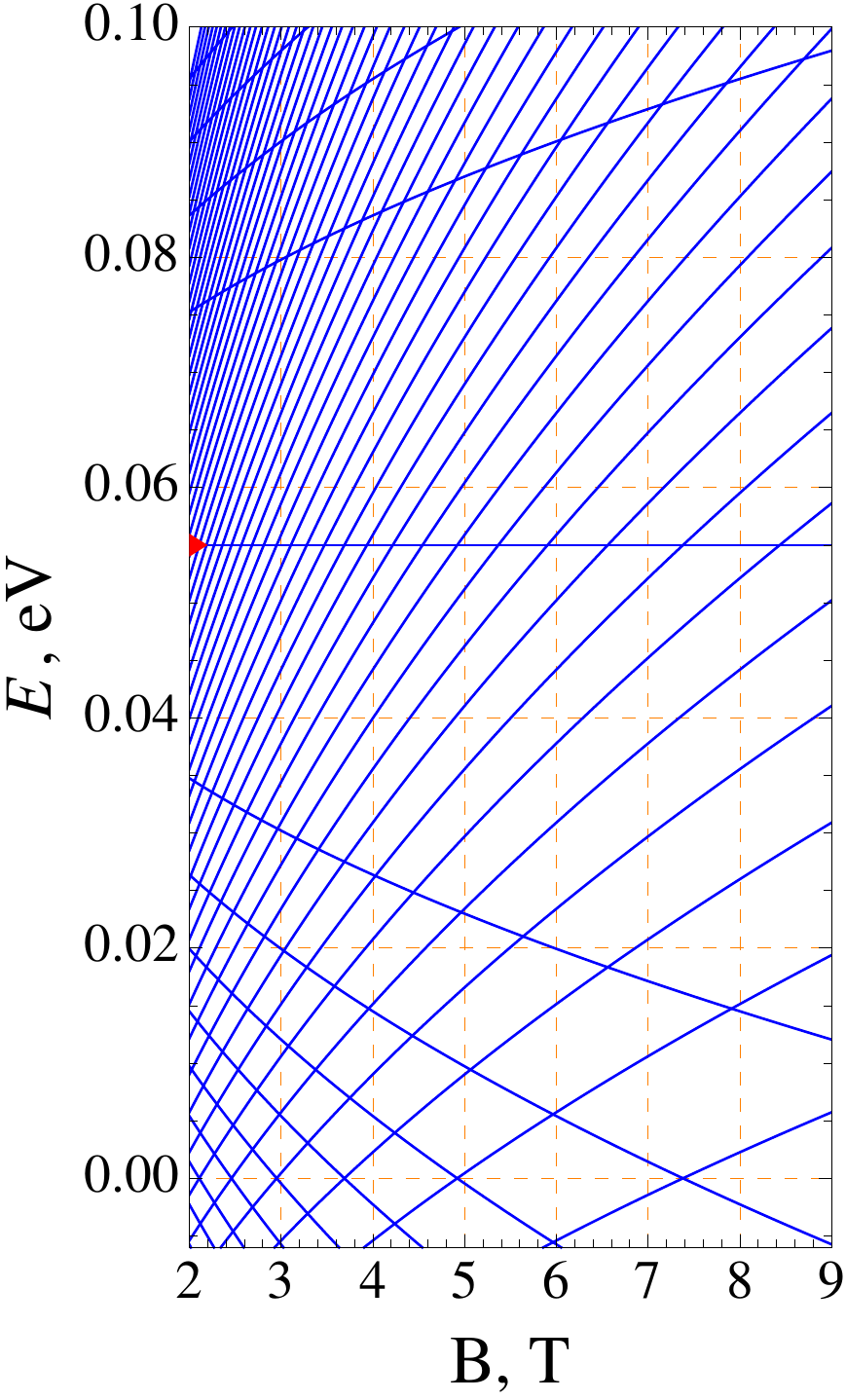}
\includegraphics[width=0.49\columnwidth]{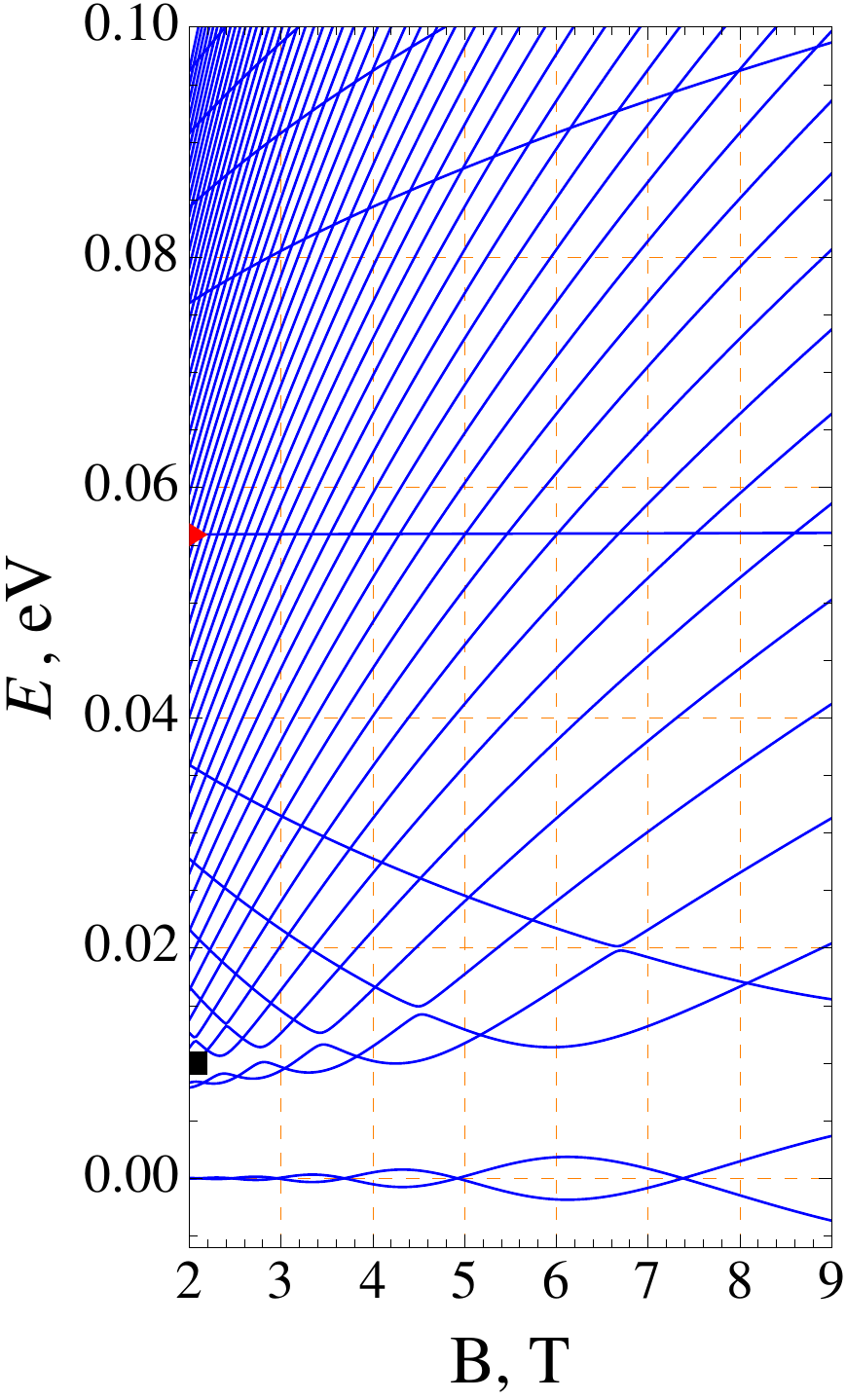}
\includegraphics[width=0.49\columnwidth]{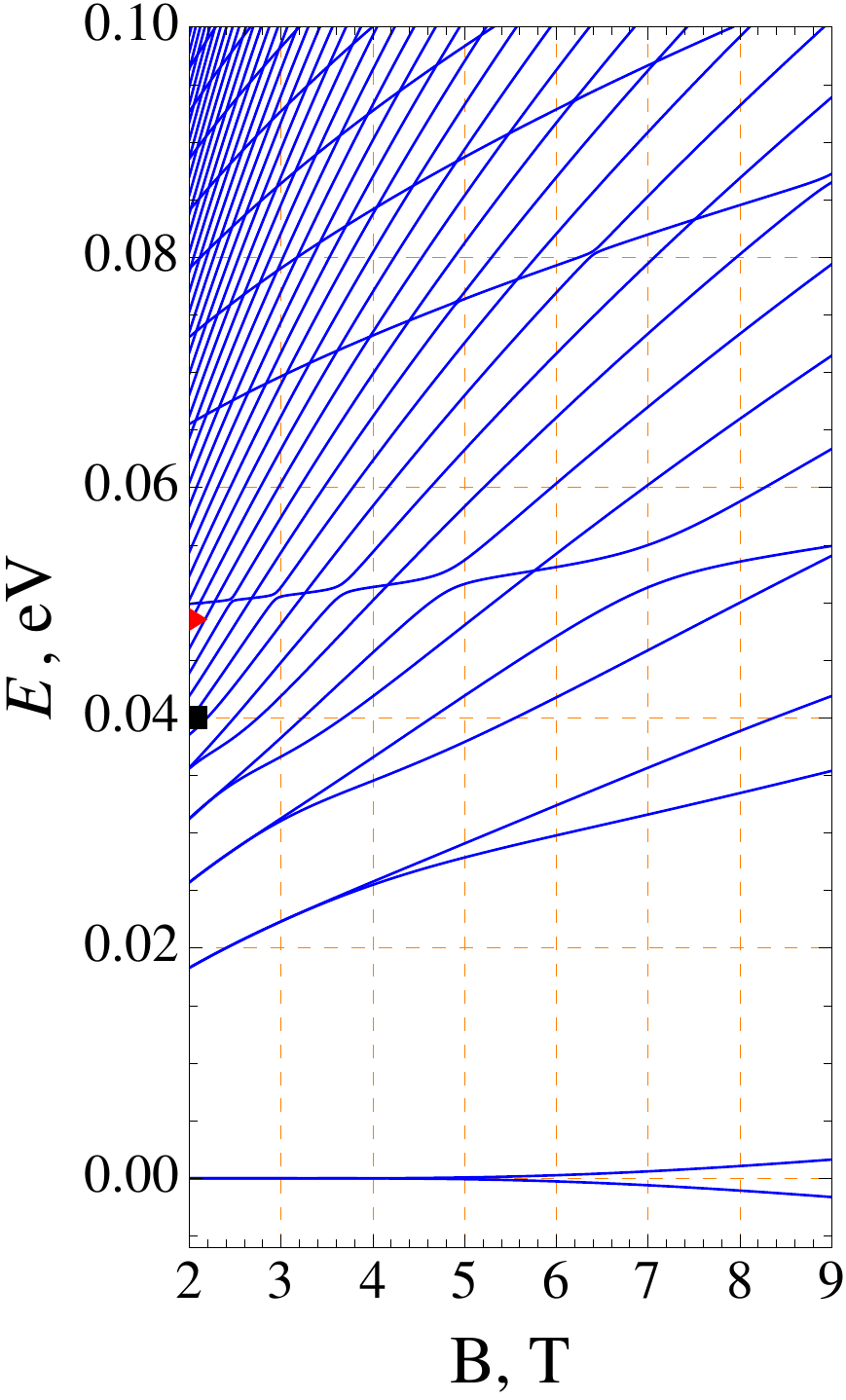}
\includegraphics[width=0.49\columnwidth]{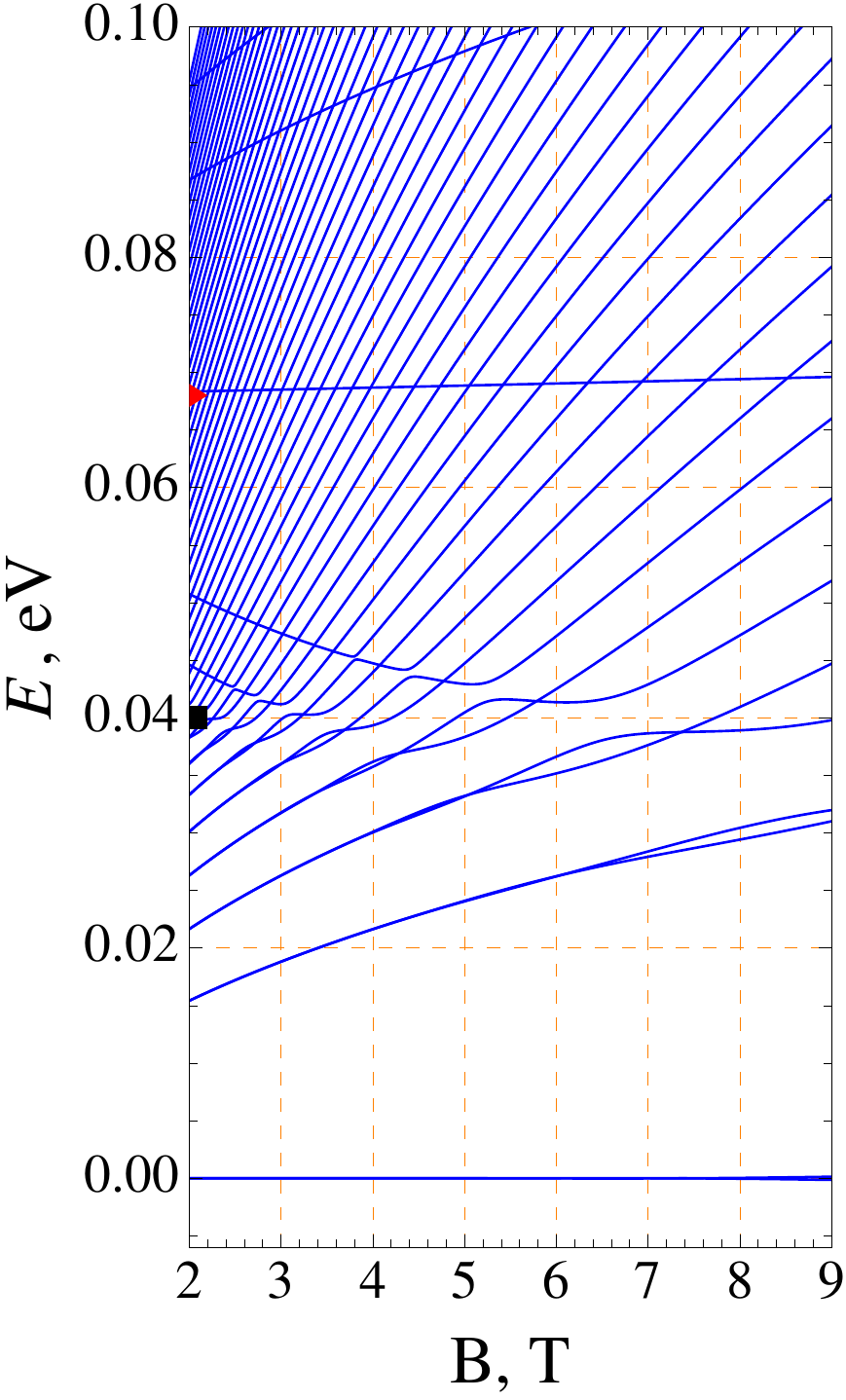}
\setlength{\unitlength}{\columnwidth}
\begin{picture}(2,0)
\put(0.0, 0.85){(a)}
\put(0.5, 0.85){(b)}
\put(1., 0.85){(c)}
\put(1.51, 0.85){(d)}
\put(0.15, 0.87){$\delta=0, m=0.055\,\text{eV}$}
\put(0.59, 0.87){$\delta=0.01\,\text{eV}, m=0.055\,\text{eV}$}
\put(1.08, 0.87){$\delta=0.04\,\text{eV}, m=0.028\,\text{eV}$}
\put(1.59, 0.87){$\delta=0.04\,\text{eV}, m=0.055\,\text{eV}$ }
\end{picture}
\caption{ \label{F:LL} Evolution of Landau levels fan diagram with variation of parameters $m$ and $\delta$ in the effective Hamiltonian. Only LL originating from $\bar X_1$ point are shown, with the vicinity of $\bar X_2$ point contributing the same set of LL.  All diagrams are symmetric around zero energy. Black square and red triangle show position of VHS, $E^\text{VH}$, and high-energy Dirac cone, $\DPht$, in the band structure. (a) For $\delta=0$, $m = 0.055\,\text{eV}$ there are two sets of Landau fans originating from two massless Dirac fermions split to lower and higher energy. (b) Small non-zero $\delta = 0.01\,\text{eV}$ in addition to $m = 0.055\,\text{eV}$ strongly changes LLs in vicinity of zero energy reflecting a formation of low-energy Dirac cones. (c) In the opposite limit $m =0.028\,\text{eV}<\delta = 0.04\,\text{eV}$ VHS strongly affects nearby 0th LL of the upper cone. (d) Fan diagram for values of $m$ and $\delta$ used in Fig.~\ref{F:band} displays well-developed LL from both low and high energy Dirac cones with a crossover happening at $E^\text{VH}$. } 
\end{center}
\end{figure*}
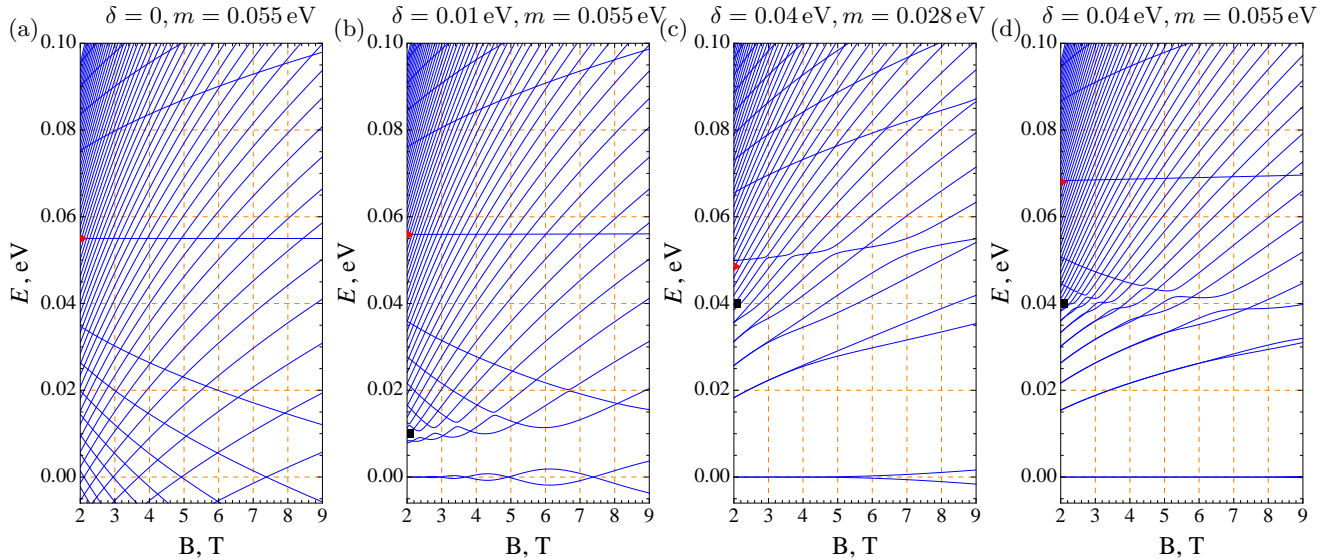

\section{Landau level spectrum \label{Sec:4}} 
Finally, we turn to the discussion of the orbital effect of the magnetic field perpendicular to the plane, which leads to the formation of Landau levels. First we aim at understanding the LL fan diagrams without mirror symmetry breaking. It has many interesting features that are unique to TCI and can be used to deduce the band structure parameters.~\cite{exp} With the understanding of the unperturbed case, we further discuss how the  various symmetry-breaking perturbations (considered in the previous Section) are manifested in the LL spectrum.

\subsection{LL fan diagram without symmetry breaking }
First, we invoke the semi-classical picture to get the basic understanding of the LL structure, which is later corroborated with the numerical calculations. Semi-classical approximation requires an integer number of magnetic flux quanta piercing the electron orbit in the real space. Relating the area of the electron orbit in the real space, to its area in the $k$-space denoted, as $S_n$, we recover the quantization condition as
\begin{equation} \label{Eq:S-semiclass}
S_n = \frac{2\pi e}{\hbar}(n+\gamma)B,
\end{equation}
where $\gamma$ is zero for Dirac fermions with linear band dispersion. Using $S(E) =  \pi E^2/{{\, {\bar v'}}}^{2}$ near the low energy Dirac point $E^{DP}_{L \pm}$ we recover the LL energy $E_n = \pm \sqrt{ 2{\bar v}' e nB/\hbar}$, where ${\bar v}' = \sqrt{v'_1 v_2} $ is the geometric mean of the Fermi velocities in two directions [see Fig.~\ref{F:band} and Eq.~(\ref{Eq:H0-proj})], and $\pm$ sign corresponds to the sign of $n$. Thus, the LL fan diagram near low energy Dirac points will consist of four-fold degenerate LL dispersing as $\sqrt{nB}$. At energies above  Van-Hove  singularity $E^\text{VH}$, the two Dirac cones merge. Thus for $E\gtrsim E^\text{VH}$ we have a sudden increase of the area of orbit in the Brillouin zone, $S(E)$, in addition to emergence of another, smaller orbit. This increase in the area $S(E)$ leads to a discontinuity 
in the LL index $n$ (based on the semiclassical scheme) at the Van-Hove singularity, and it was indeed observed in recent experiment.~\cite{exp} More quantitatively, the degeneracy between two  LL levels with index $n$ is lifted and one gets a LL with index $2 n+ k$ associated with the larger outer Fermi pocket and another LL with index $-k$ ($k \geq 0$) associated with the smaller inner Fermi hole pocket.

Further above Van-Hove singularity, the band structure is again well described by two high-energy Dirac cones, both having the same mean Fermi velocity $\bar v = \sqrt{v_1 v_2}$ but displaced in energy. Thus it is qualitatively expected to look like a dense sequence of LLs from the Dirac cone with bigger area, pierced by a sparsely separated LL from the interior Dirac cone. 

To reveal additional features beyond the semi-classical approach, we numerically calculate the LL spectrum of the pristine TCI (001) surface states using the four-band Hamiltonian~(\ref{Eq:Hsurf}). This is achieved by replacing operators $v_2 k_2 \pm i v_1 k_1$ in Eq.~(\ref{Eq:Hsurf}) by ladder operators, $\pi, \pi^\dagger$ acting in the basis of Landau level orbitals with matrix elements:
\begin{subequations}\label{Eq:pi}
\begin{eqnarray} 
 \pi | n  \rangle & =& \frac{\hbar \bar v }{\ell_B} \sqrt{2n} | n-1  \rangle,
\\
 \pi^\dagger | n  \rangle & =& \frac{ \hbar  \bar v }{\ell_B} \sqrt{2(n+1)} | n+1  \rangle,
\end{eqnarray}
\end{subequations}
where the magnetic length $\ell_B = \sqrt{\hbar/(eB)}$, and $\bar v  = \sqrt{v_1 v_2}$.  In the basis of Landau orbitals and valley/spin degrees of freedom,  
the Hamiltonian in the presence of a magnetic field becomes a $4 \Ld \times 4\Ld$ matrix, where we impose a cutoff $\Ld$  corresponding to the highest Landau orbital ($\Ld=100$ in all plots presented here). The LL spectrum is given by the eigenvalues of this matrix. Details of the calculation can be found in, for example, Ref.~\onlinecite{SerbynABA}

We plot the LL fan diagram from 2T to 9T in Fig.3, for several values of band structure parameters $m$ and $\delta$. 
All cases show two different sets of LLs, associated with emergent low energy Dirac cones, and those from energies above VHS, as expected from the 
semiclassical analysis. However, many important features of the LL spectrum depends on $m$ and $\delta$.

The LL fan diagram of the Hamiltonian~(\ref{Eq:Hsurf}) with $\delta = 0$ in Fig.~\ref{F:LL}(a) displays two sets of LLs varying as $\sqrt{B}$ with magnetic field~(Figure~\ref{F:LL} shows LL with $E>0$, as all LL fan diagrams are symmetric around zero energy). This is in full agreement with the band structure for $\delta = 0$, given by the two Dirac cones split in energy by $\pm m$. VHS is absent in this case, and there are only two non-dispersive LL which are 0th LL of corresponding Dirac cones $\DPhot$.   Non-zero $\delta$ leads to appearance of two emergent low-energy Dirac cones, however the energy range where such description is restricted to be below VHS, $|E|<\delta$. Indeed, in Fig.~\ref{F:LL}(b) for small magnetic fields we see the formation of non-dispersive doubly degenerate 0th LL associated with low-energy Dirac cones. For stronger magnetic fields, when $\Delta k \ell_B \sim 1$, where $\Delta k$ is the distance between the origin of two low energy Dirac cones, the 0th LL is split and the splitting oscillates with magnetic field, which is a consequence of magnetic breakdown.  The 1st LL is also visible in  Fig.~\ref{F:LL}(b), though it is located very close to VHS and thus does not follow $\sqrt{B}$ dependence well. Also, $\Delta k$, defined now as a distance between two Fermi surfaces in the momentum space,  becomes smaller as we approach VHS, thus the magnetic breakdown happens for weaker magnetic fields. 

The opposite limit of $\delta$ larger than $m$ in Fig~\ref{F:LL}(c) has well-developed 0th LL and three higher LLs of the low energy Dirac cones. These LLs are doubly degenerate when the magnetic field is not strong enough and $1/\ell_B$ is smaller than the distance between different Fermi surfaces. Note, that 0th LLs associated with the $\DPhot$ Dirac cone are also affected by the nearby VHS: it is the same magnetic breakdown which leads to a series of avoided crossings between the 0th LL and other LL at the same energy.

\begin{figure}[b]
\begin{minipage}[c]{0.99\columnwidth}
\includegraphics[width=0.49\columnwidth]{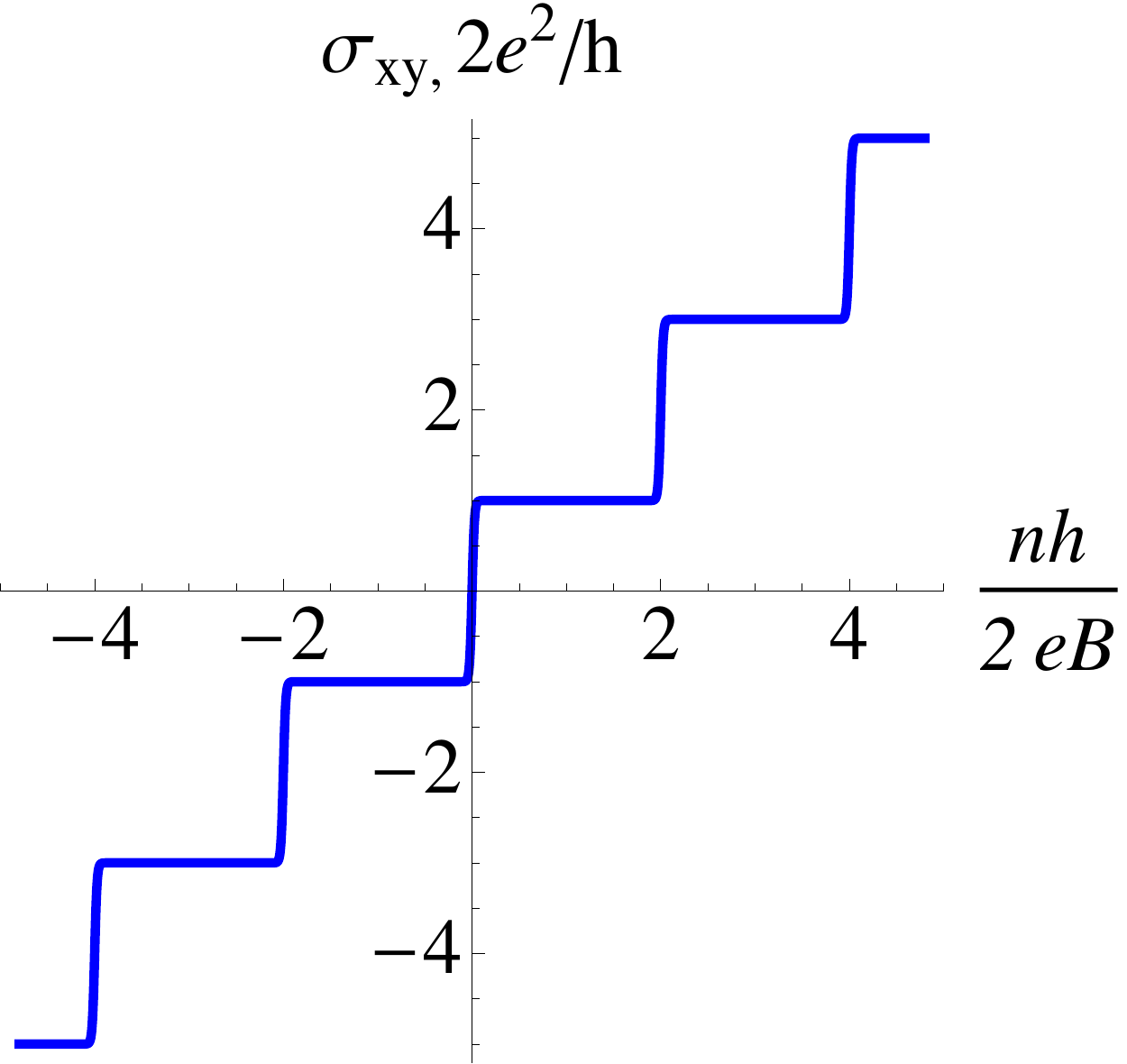}
\includegraphics[width=0.49\columnwidth]{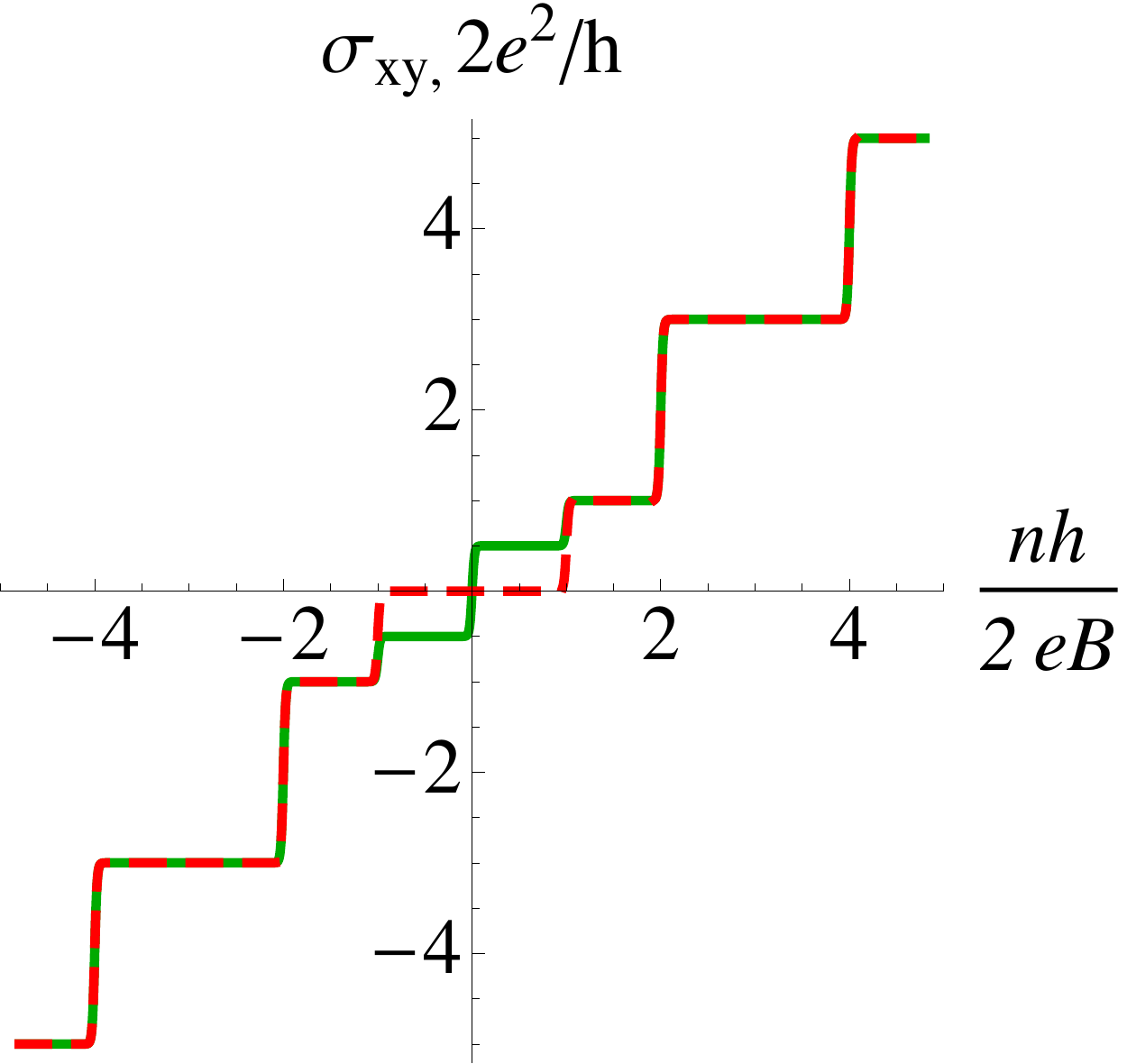}
\end{minipage}%
\hfill
\setlength{\unitlength}{\columnwidth}
\begin{picture}(1,0)
\put(0.0, 0.4){(a)}
\put(0.45, 0.4){(b)}
\end{picture}
\caption{ \label{F:LLsigma} (a) QHE plateaus from low energy Dirac cones when there is no symmetry breaking. (b) QHE plateaus when there is a $M_1$ breaking (solid blue line) and both $M_1$ and $M_2$ are broken (dashed red line).}
\end{figure}
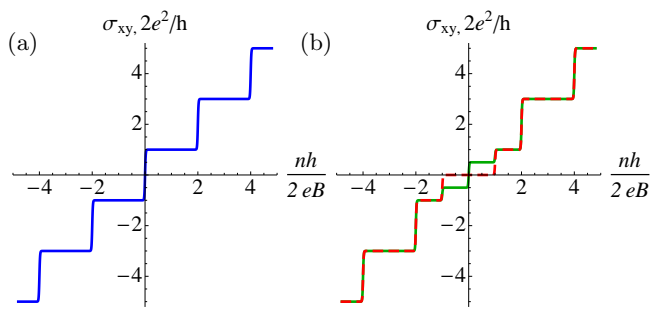

Finally, Fig.~\ref{F:LL}(d) presents LL fan diagram for parameters $m = 0.055\,\text{eV}$, $\delta = 0.04\,\text{eV}$. These LL were recently observed in LL STM spectroscopy experiment in Ref.~\onlinecite{exp} and were used to determine the values of parameters in the effective Hamiltonian. When $m$ is comparable to $\delta$ there is a series of well-resolved doubly degenerate LL from low-energy Dirac cones~[Eq.~(\ref{Eq:H0-proj})]. For energies above VHS these LL cross over into singly-degenerate LL well approximated by $\DPhot$ Dirac cones.

The particular degeneracy pattern of LL arising when $m$ and $\delta$ are of the same magnitude should be also visible in transport measurements. For this we have to recall that there are two points $\bar X_{1,2}$ in the BZ with the similar band structure. Thus, when there is no symmetry breaking, LLs are \emph{four-fold} degenerate at lower energy and \emph{two-fold} degenerate at higher energies. This should give rise to the sequence of QHE plateaus with 
\begin{equation} \label{Eq:sxy-low}
  \sigma^\text{l-e}_{xy} 
  = 
  \frac{4e^2}{h} (n+1/2)
\end{equation}
in vicinity of neutrality point [see Fig.~\ref{F:LLsigma}(a)]. Notable distinction with graphene~\cite{Novoselov06} is that here the factor of four arises from the presence of two $\bar X$ points and two Dirac cones emergent in vicinity of each $\bar X$ point, rather than from valley and spin degeneracy. For higher filling factors (or at higher magnetic fields), when the two-fold degeneracy from two copies of low energy Dirac cones is lifted, the $\Delta \sigma_{xy}$ between adjacent plateaus becomes twice smaller,
\begin{equation} \label{Eq:sxy-high}
  \Delta \sigma^\text{h-e}_{xy} 
  = 
  \frac{2e^2}{h}.
\end{equation}

\subsection{Consequences of symmetry breaking  for LL}
Following our discussion of  mirror symmetry breaking effect for the band structure, we study its manifestation for the LL spectra and transport measurements.

\begin{figure*}
\includegraphics[width=0.49\columnwidth]{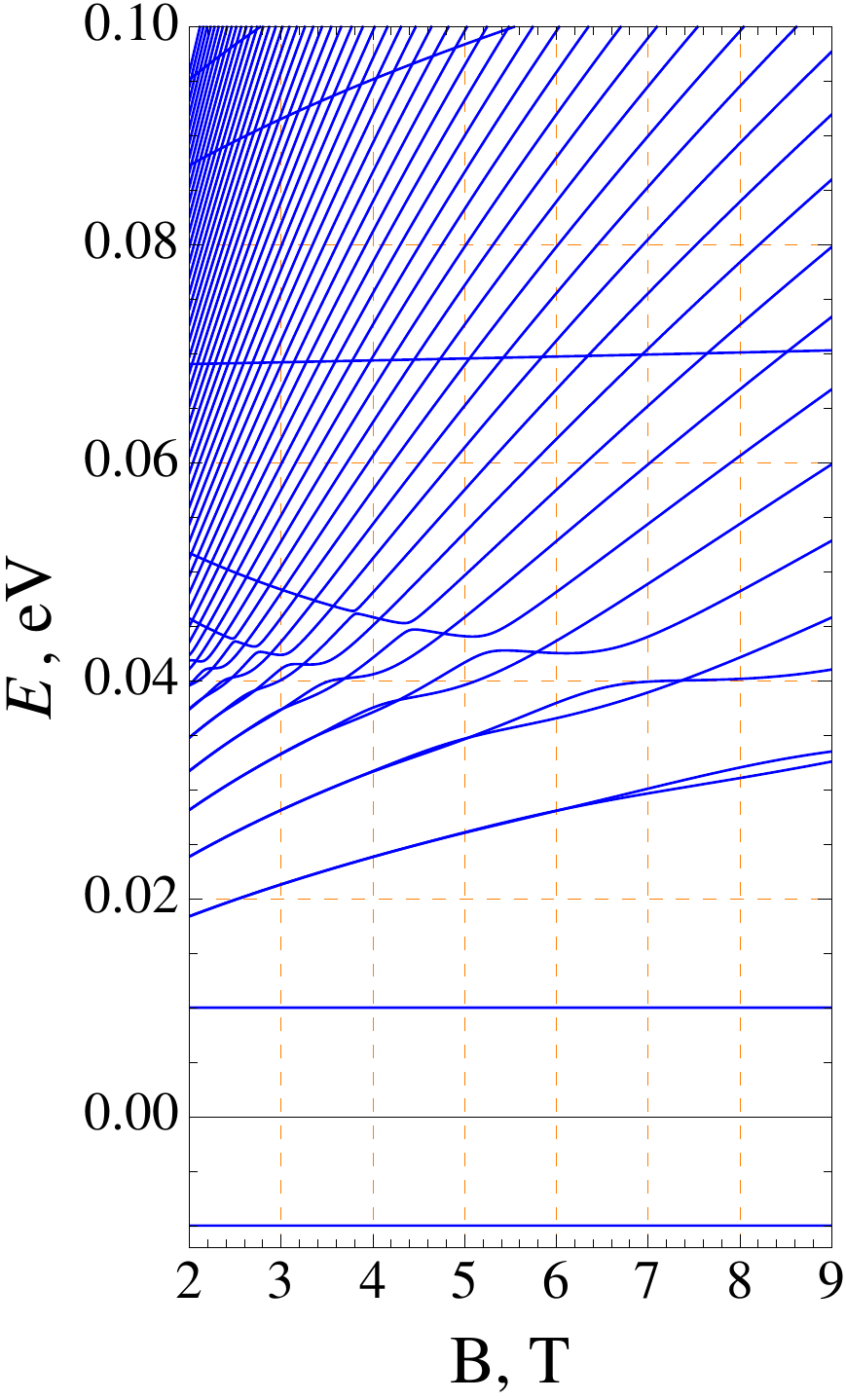}
\includegraphics[width=0.49\columnwidth]{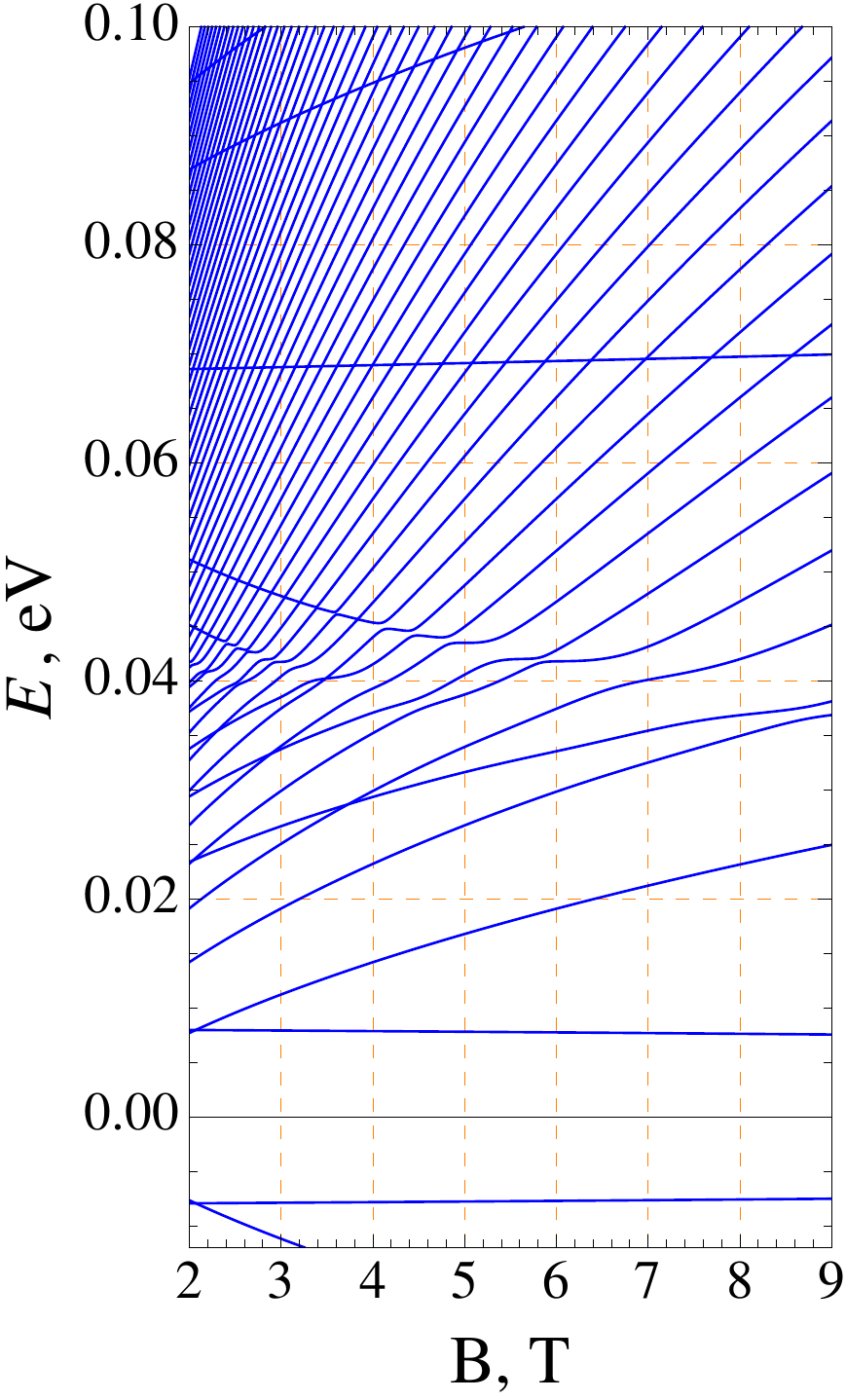}
\includegraphics[width=0.49\columnwidth]{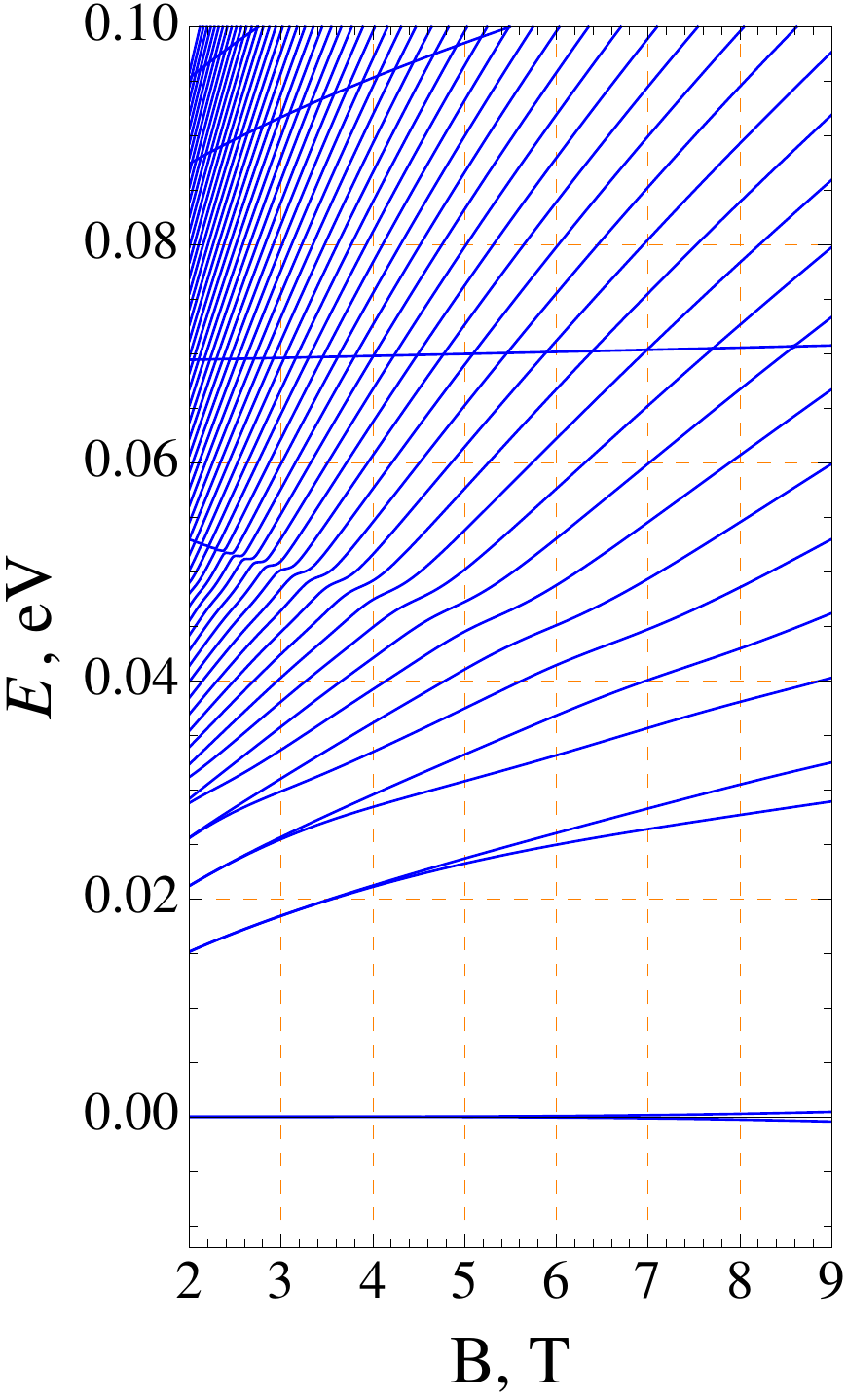}
\includegraphics[width=0.49\columnwidth]{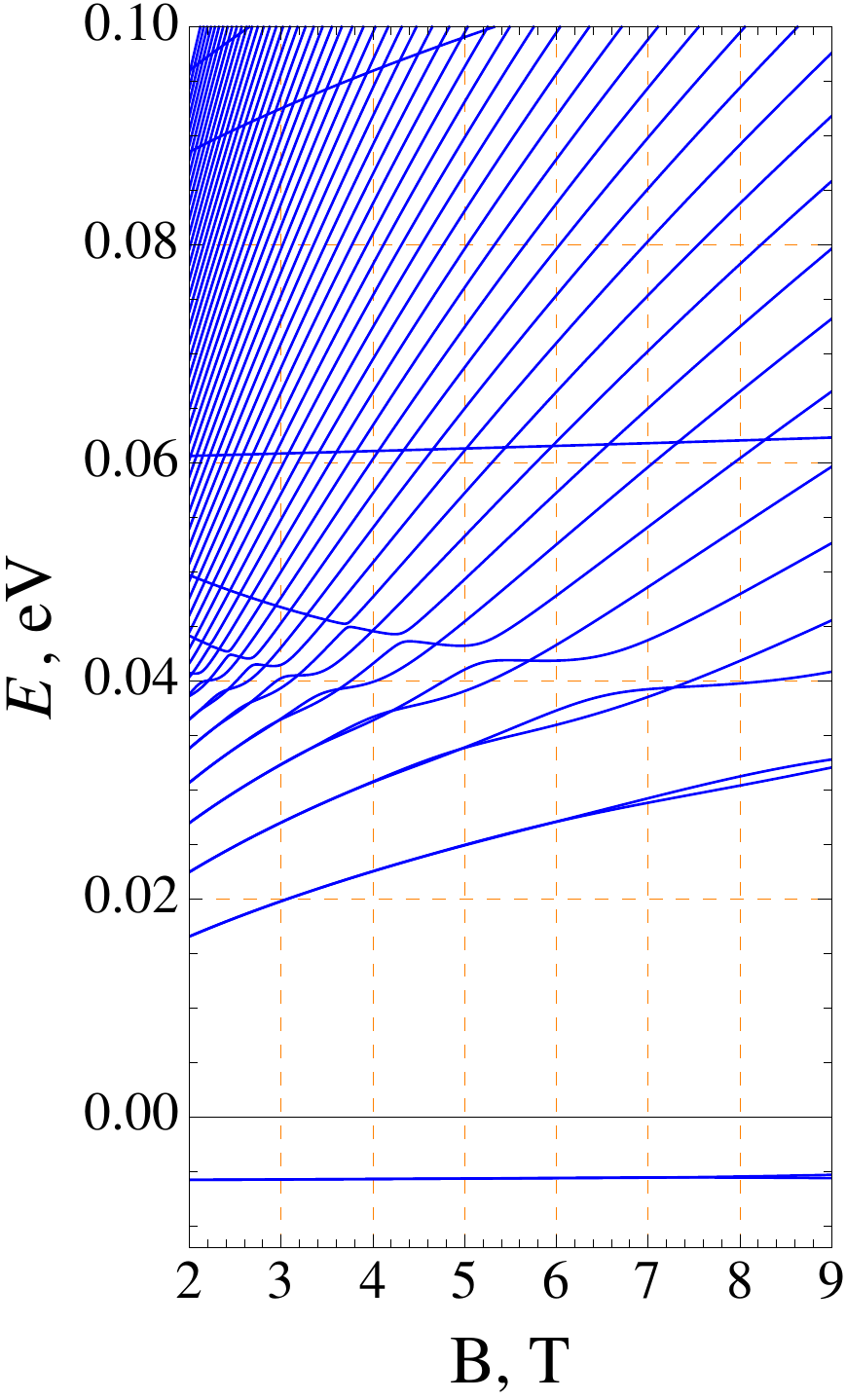}
\setlength{\unitlength}{\columnwidth}
\begin{picture}(2,0)
\put(0.0, 0.85){(a)}
\put(0.5, 0.85){(b)}
\put(1., 0.85){(c)}
\put(1.51, 0.85){(d)}
\put(0.2, 0.87){$\delta H_{\bar X_1} = \Delta_F \tau_z$}
\put(0.65, 0.87){$\delta H_{\bar X_1} = \lambda_1 B_1 s_x\tau_x$}
\put(1.15, 0.87){$\delta H_{\bar X_1} =\lambda_2  B_2 s_z \tau_z$ }
\put(1.65, 0.87){$\delta H_{\bar X_1} =\mu^B_3 B_3 s_z$ }
\end{picture}
\caption{ \label{F:LLmass} LL from the band structure in vicinity of $\bar X_1$ when the symmetry breaking perturbation is present. (a) The main effect of the ferroelectric distortion is the splitting of low energy 0th LL.  (b) Application of  in-plane magnetic field along $(100)$ direction shifts LL fans corresponding to $\DPpm$ up and down in energy also splitting 0th LL and resulting in additional multiple level crossings. (c) Effect of the non-zero component $B_2$ of in-plane magnetic field. Perturbation mostly affects the onset of LL splitting and vicinity of Van-Hove singularity. (d)   Zeeman-like coupling to the $B_3$ component of magnetic field  shifts 0th LL of $\DPpm$  to negative energy, breaking the particle-hole symmetry.  The strength of perturbation in all cases chosen as $\Delta_F = \lambda_1 B_1 = \lambda_2 B_2  = \mu^B_3B_3 =  10\,\text{meV}$.}
\end{figure*}
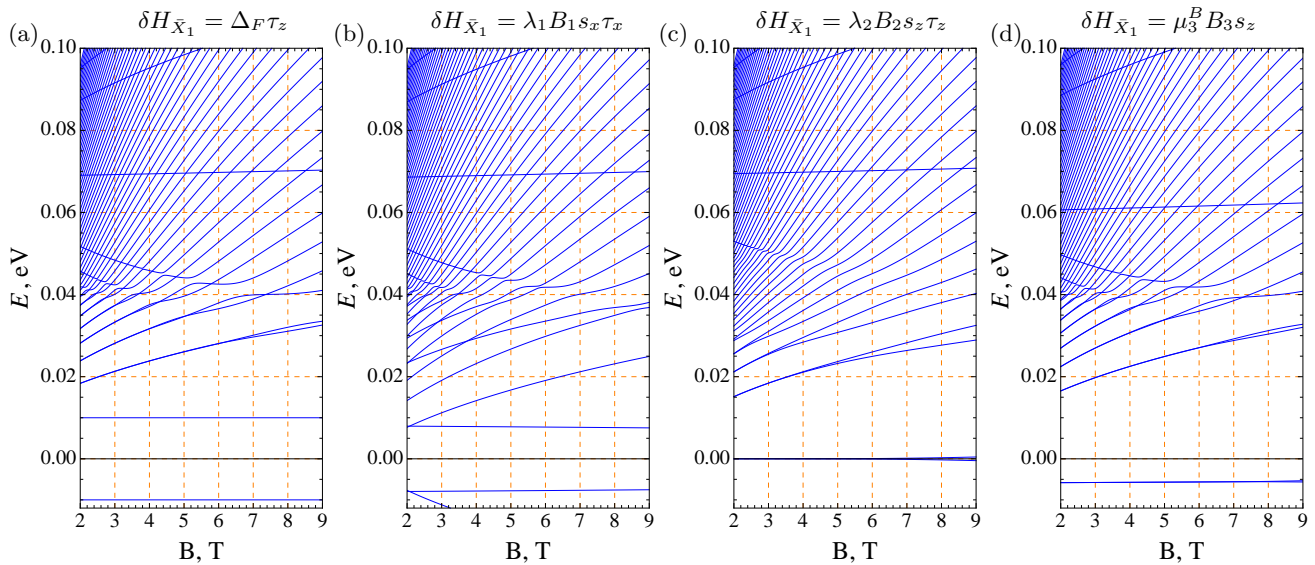

\emph{Orthorhombic distortion} which breaks only one of two mirror planes, $M_1$ for concreteness, gaps out both low energy Dirac cones near $\bar X_1$ point [see Eq.~(\ref{Eq:HM1proj})]. The masses have opposite sign for $\DPpm$ points, as dictated by unbroken time-reversal symmetry. On the other hand, the  band structure in vicinity of $\bar X_2$ is weakly affected by the breaking of $M_1$ mirror. Thus, in magnetic field,  the four-fold degeneracy of low energy 0th LL will be partially lifted: in vicinity of $\bar X_1$ point, the 0th LL will be split from zero energy to $\pm \Delta$~[see Fig.~\ref{F:LLmass}(a)]. The Dirac fermions in vicinity of $\bar X_2$ point will remain massless so that the LL structures shown in Fig.~\ref{F:LLmass}(a) and  \ref{F:LL}(d) will coexist. This results in a peculiar structure of doubly degenerate LL at zero energy surrounded by two singly-degenerate 0th LLs at $\pm \Delta$. The emergent pattern of plateaus in  QHE is shown in Fig.~\ref{F:LLsigma}(b)~[solid blue line]: the  height of the step in $\sigma_{xy}$  at the neutrality point is now  $\Delta \sigma_{xy} = 2 e^2/h$, being two times smaller than in the unbroken symmetry case. However, there is new plateau with $\Delta \sigma_{xy} = e^2/h$ due to the split LL. Note, that the value of the splitting observed experimentally, $\Delta \approx 10\,\text{meV}$ should allow for resolving this additional plateau at low temperatures. 

One cannot exclude the possibility of \emph{orthorhombic distortion breaking both} $M_1$ and $M_2$ mirror symmetries, thus gapping out Dirac fermions in vicinity of both $\bar X_{1,2}$ points and fully splitting 0th LL. In transport this will manifest itself as appearance of plateau at the neutrality point, see dashed line in Fig.~\ref{F:LLsigma}(b). Observation of such symmetry breaking opens interesting possibility of realizing domain walls between different regions where the $\Delta$ controlling the symmetry breaking strength has different sign.  Without magnetic field, two-dimensional Dirac fermions with mass $\Delta_{\bar X_1}$ changing sign will have one-dimensional zero-energy modes localized near such domain wall. More specifically, to maintain the time reversal symmetry, a pair of counter-propagating edge states protected by Kramers degeneracy should arise. In magnetic field, the 0th LL split from zero would bend towards zero energy, restoring the four-fold degeneracy in vicinity of the domain wall. Thus, such domain walls may be visible in the spatially resolved LL spectroscopy on STM. 

\emph{Strain}, as was argued in Section~\ref{Sec32}, modifies the band structure parameters $m$ and $\delta$, and can shift $\DPpm$ points away from $k_1=0$ line. 
These effects leads to the modification of position of VHS, shift of the $\DPhot$ and change of the onset of magnetic breakdown, which can be detected by the LL  spectroscopy. 

\emph{In-plane magnetic field} can shift the position of the low-energy Dirac cones in the BZ, but this is not readily observable. In addition, $B_1$ component of magnetic field  induces asymmetry between $\DPpm$ points located near $\bar X_1$ point. The resulting modification of the LL fan diagram is shown in Fig.~\ref{F:LLmass}(b). On the other hand, as we discussed above,  the effect of the $B_1$ on the vicinity of $\bar X_2$ point can be understood from the effect of $B_2$ near $\bar X_1$ point. The latter is illustrated in Figure~\ref{F:LLmass}(c). Therefore, the full LL fan diagram accounting for vicinity of $\bar X_{1,2}$ points consists of LLs shown in Fig~\ref{F:LLmass} in panels (c) and (d).

\emph{Out-of-plane magnetic field} leads to appearance of  Landau levels.  On the other hand, \emph{magnetic impurities} can induce Zeeman-type effects. Zeeman-type couplings in Eq.~(\ref{Eq:B3})  gap low-energy Dirac cones with the mass of the same sign. In particular,   Fig.~\ref{F:LLmass}(d) illustrates of the effect of $\mu_{B_3} B_3 s_z$ on LLs.   Note, that account for the  contribution from the $\bar X_2$ point leads to a two-fold increase in the degeneracy of all LL in Fig.~\ref{F:LLmass}(d).

\section{Summary\label{Sec:5}}
In summary, we studied the effect of various symmetry breaking perturbations on the surface band structure within an effective model. All perturbations considered by us are potentially realizable: ferroelectric distortion naturally occurs in IV-VI semiconductors. Strain can be applied in a controlled manner with existing experimental techniques. Finally, doping with magnetic impurities and (or) application of magnetic field can realize time-reversal breaking perturbations. 

By supplementing the effective $\bk\cdot \bp$ model~\cite{JunweiLiu13} with symmetry breaking terms derived here, we have deduced the effects of symmetry breakings on electronic properties of TCI surface states, and described their experimental signatures. We have found that many types of symmetry breaking perturbations leave distinctive fingerprints in the Landau level spectrum of TCI surface states, some of which have been observed by STM.~\cite{exp} We have predicted magneto-transport properties of TCI surface states in the presence of symmetry breakings.

\section*{Acknowledgments}
We thank V. Madhavan and Y. Okada for related collaborations, and P. A. Lee for discussions. M.S. was supported by P. A. Lee via grant NSF DMR 1104498. L. F. is supported by the DOE Office of Basic Energy Sciences, Division of Materials Sciences and Engineering under award DE-SC0010526

\end{document}